\documentclass[twocolumn]{aastex63}

\begin{document}
\title{The Galactic Chemical Evolution of Chlorine}

\author{Z. G. Maas} 
\affil{University of Texas at Austin, McDonald Observatory, Austin, TX 78712, USA}

\author{C. A. Pilachowski}
\affil{Indiana University, Bloomington, IN 47405, USA}

\email{zgmaas@utexas.edu}

\begin{abstract}

We measured $^{35}$Cl abundances in 52 M giants with metallicities between --0.5 $<$ [Fe/H] $<$ 0.12. Abundances and atmospheric parameters were derived using infrared spectra from CSHELL on the IRTF and from optical echelle spectra. We measured Cl abundances by fitting a H$^{35}$Cl molecular feature at 3.6985 $\mu$m with synthetic spectra. We also measured the abundances of O, Ca, Ti, and Fe using atomic absorption lines. We find that the [Cl/Fe] ratio for our stars agrees with chemical evolution models of Cl and the [Cl/Ca] ratio is broadly consistent with the solar ratio over our metallicity range. Both indicate that Cl is primarily made in core-collapse supernovae with some contributions from Type Ia SN.  We suggest other potential nucleosynthesis processes, such as the $\nu$-process, are not significant producers of Cl. Finally, we also find our Cl abundances are consistent with H II and planetary nebular abundances at a given oxygen abundance, although there is scatter in the data. 

\end{abstract}


\section{Introduction}
\label{sec::intro}

The odd-Z element chlorine has few chemical abundance measurements in stars, yet is an intriguing element due to the multiple possible ways both stable isotopes of Cl, $^{35}$Cl and $^{37}$Cl, are thought to be produced. Stellar chlorine abundances are difficult to measure since no detectable optical or infrared atomic Cl lines have been found in FGK stars. Instead, HCl molecular lines have been used to measure Cl abundances. Due to the low Cl abundance relative to the even-Z elements and low dissociation energy of the HCl molecular \citep{martin98}, only the coolest stars have detectable HCl absorption lines. For example, $^{35}$Cl abundances have been measured in the sun via sunspot umbrae spectra \citep{hall72} and in 16 M stars \citep{maas16}.  

The Cl isotope ratio may be measured in molecular clouds using Cl bearing molecules and Cl nebular abundances are measured from forbidden Cl transitions in planetary nebulae and H II regions. H II region abundances have compared Cl to O and found the [Cl/O] ratio is nearly constant with galactocentric radius and therefore both elements evolve in lockstep \citep{esteban15,arellano20}. H II regions in M31 also show a relationship between Cl and O with more scatter \citep{fang18}. Similar Cl to O correlations have been found in planetary nebulae in the Galaxy \citep{henry04,milingo10} and abundances have shown that Cl evolves in lockstep with Ar \citep{delgado15}. 

The Cl isotope ratio has varies both in the interstellar medium (ISM) and in a small sample of measured M giants. $^{35}$Cl and $^{37}$Cl bearing molecules have been measured in the circumstellar envelopes of evolved stars such as IRC+10216 \citep{cernicharo00,kahane00,highberger03}, in molecular gas in star formation regions such as the Orion Nebula \citep{salez96, cernicharo10, peng10, neufeld12, neufeld15, kama15}, towards bright infrared sources, and in intergalactic sources, along sight-lines toward quasars \citep{muller14,wallstrom19}. Combining the results from all the surveys reveals a large range in Cl isotope ratios from 1 $<$ $^{35}$Cl/$^{37}$Cl $<$ 5 in the ISM. A scatter of 1.76 $<$ $^{35}$Cl/$^{37}$Cl $<$ 3.42 was also measured in a small sample of six M giants \citep{maas18}. The origin of the Cl isotope ratio scatter is still unknown but may be due to differences in local enrichment or self-enrichment. For example, the s-process may lower the Cl isotope ratio when compared to the solar value \citep{highberger03,cristallo15,karakas16}.

Both Cl isotopes are predicted to be produced in core collapse supernovae (CCSNe) during explosive oxygen with $^{35}$Cl produced from proton capture on sulfur and $^{37}$Cl from the radioactive decay of $^{37}$Ar \citep{woosley73,thielemann85,woosley95}. Both isotopes may be produced in Type Ia supernovae, however the contribution is not thought to be as significant as CCSNe \citep{travaglio04,leung18}. The weak s-process in massive stars is thought to be a significant source of $^{37}$Cl \citep{prantzos90,pignatari10} while the s-process in AGB stars is not thought to be as important \citep{cristallo15,karakas16}. $^{35}$Cl may also be produced from neutrino spallation during CCSNe \citep{woosley90}. Finally, the interaction and merger of the carbon and oxygen burning shells in massive stars may increase the production of the odd-Z light elements, including Cl \citep{ritter18}. A summary of the yields from the different proposed nucleosynthesis sites of Cl are listed in Table 1 of \citet{maas18}. 

Chemical evolution models predict the Cl to iron ratio will be nearly constant between --0.16 $\lesssim$ [Cl/Fe] $\lesssim$ --0.26 over --4 $<$ [Fe/H] $<$ 0.2 \citep{kobayashi11,prantzos18}. Contributions from rotating massive stars are important at low [Fe/H] although less significant between --1 $<$ [Fe/H] $<$ 0.2 \citep{prantzos18}. Other chemical evolution models suggest a more complex evolution of Cl, with [Cl/Fe] $\sim$ --0.8 at [Fe/H] = --2 \citep{kobayashi20}. A sample of 16 stars with Cl abundances between --0.72 $<$ [Fe/H] $<$ 0.2 was broadly consistent with chemical evolution models and possibly models under-predicted the measurements \citep{maas16}. However, the sample size was small and the typical uncertainties were 0.23 dex. 

To understand the production sites and Galactic chemical evolution of chlorine, we have measured Cl abundances in 52 M giant stars, as discussed in \citet{maas20}. Section \ref{sec:obs_reduction} describes the observations and data reduction. Section \ref{sec::atmo_params} details the atmospheric parameter ($T_{\mathrm{eff}}$, log(g), [Fe/H], microturbulence) derivations. Elemental abundance and uncertainty measurements are characterized in section \ref{sec:abun_measure}. The results are discussed in section \ref{sec:discussion} and the conclusions summarized in section \ref{sec:conclusion}.

\section{Observations and Data Reduction}
\label{sec:obs_reduction}
\subsection{Sample Selection}

We chose sample stars first based on their effective temperature and their observed K$_{s}$ magnitude. The HCl molecule has a low dissociation energy \citep{martin98} and therefore only low temperature stars are able to form detectable HCl ro-vibrational features. \citet{maas16} examined stars from a range of $T_{\mathrm{eff}}$ from 3300 K to 4300 K and found only stars with $T_{\mathrm{eff}}$ $\lesssim$ 4000 K had detectable H$^{35}$Cl features. The sky background is also significant at 3.7 $\mu$m, the approximate location of the targeted HCl molecular features, and can dominate the observed signal for long exposure times. Therefore, to achieve the signal-to-noise ratios necessary (S/N $>$ 100 from \citealt{maas16}), stars with K$_{s}$ $<$ 4 mag were chosen for observations using DIRBE \citep{hauser98,smith04} and 2MASS photometry \citep{skrutskie06}.

We also derived thin and thick disk membership probabilities for the stars meeting both criteria listed above. The population membership probabilities were derived using U V W Galactic velocity components from \citet{famaey05} and equations 1 and 2 from \citet{ramirez13}. Both thin and thick disk stars were selected to explore the abundance pattern of the disk. Finally, binaries and variable stars were removed from the sample using the SIMBAD database\footnote{\url{http://simbad.u-strasbg.fr/simbad/}}. 

Few M giants in the sample had previously determined atmospheric parameters in the literature. Optical echelle spectra for each target star was obtained, along with IR spectra, in order to determine [Fe/H] abundances, microturbulence values, and abundances of additional elements. We obtained high resolution spectra for 52 stars in total; a list of the final sample is given in Table \ref{table::obslog}. 

\begin{deluxetable}{ c c c c c c}
\tablewidth{0pt} 
\tabletypesize{\footnotesize}
\tablecaption{Sample and Observation Summary \label{table::obslog}} 
 \tablehead{\colhead{HD} & \colhead{K$_{s}$} & \colhead{K$_{s}$} & \colhead{S/N} & \colhead{Echelle} & \colhead{S/N}\\
 \colhead{Num.} & \colhead{(Mag.)} & \colhead{Source}& \colhead{(IR)} & \colhead{Source} &  \colhead{(Optical)}}
\startdata
5111	& 	2.8	& 1 &	70	&	1	&	40	\\
15594	&	1.3	& 1 &	100	&	1	&	120	\\
19258	&	2.14 & 2 	&	150	&	1	&	60	\\
25000	&	1.53 & 2 	&	130	&	1	&	90	\\
30634	&	2.35 & 2 	&	200	&	1	&	60	\\
31072	&	3.05 & 2 	&	130	&	1	&	60	\\
35067	&	3.08 & 2 	&	50	&	1	&	50	\\
84181	&	3.0 & 1	&	150	&	1	&	90	\\
99592	&	0.52& 2 	&	180	&	1	&	100	\\
102159	&	0.78& 2 	&	160	&	2	&	20	\\
103340	&	3.9 & 1	&	40	&	1	&	50	\\
108815	&	1.87 & 2 	&	200	&	1	&	80	\\
114941	&	3.54 & 2 	&	110	&	1	&	80	\\
115322	&	1.36 & 2 	&	150	&	1	&	110	\\
\enddata
\tablecomments{K$_{s}$ Sources: (1) 2MASS \citep{skrutskie06}, (2) DIRBE \citep{hauser98,smith04}}
\tablecomments{Echelle Spectra Sources: (1) ARCES on APO 3.5m, (2) TOU on 50" Dharma Endowment Foundation Telescope}
\tablecomments{(This table is available in its entirety in machine-readable form.)}
\end{deluxetable}

\subsection{IRTF Observations and IR Spectra Data Reduction}

Our target stars were observed with CSHELL \citep{tokunaga90,greene93} on the 3m NASA Infrared Telescope Facility (IRTF). The central wavelength for observations was 36980 $\mbox{\AA}$ to best observe the H$^{35}$Cl feature at 36985.1 $\mbox{\AA}$ and the total wavelength range is $\sim$ 70 $\mbox{\AA}$. A slit width of 0.5" was chosen to achieve a resolution of R $\sim$ 40,000, necessary to resolve the H$^{35}$Cl feature from a nearby OH line.

Data reduction was performed with PyRAF\footnote{PyRAF is a product of the Space Telescope Science Institute, which is operated by AURA for NASA.} following standard infrared spectroscopic data reduction procedures \citep{joyce92}. First, the raw observations were trimmed, flatfielded, and dark subtracted using median combined calibration images. Targets were nodded along the slit and pairs of images were taken in two different spatial positions. We then subtracted the nodded pair of images from one another to remove the bias and sky background. Spectra were then extracted from each observation. A wavelength solution derived from atomic Ca I and OH lines were applied to the data using the linelist in \citet{maas16}. The spectral range contains too few telluric features for an independent wavelength solution. Signal-to-noise calculations are given in Table \ref{table::obslog} and were estimated by calculating the standard deviation on observed data from the normalized continuum in regions with no stellar absorption features. The specific regions used are between 36948 $\mbox{\AA}$ - 36952 $\mbox{\AA}$, 36958 $\mbox{\AA}$ - 36961.1 $\mbox{\AA}$, and from 37013 $\mbox{\AA}$ - 37019 $\mbox{\AA}$.

After the wavelength solution was calculated for extracted stellar spectra, they were median combined. Any observations with apertures near the edge of the detector when nodded were not used in the final combination to ensure only accurately traced images were included. After the wavelength solution was applied, telluric lines were divided out of the object spectrum using observations of spectral type A and B telluric standard stars. The airmass difference between the telluric standard and object was limited to 0.1 and the airmass was $\leq$ 1.1 for most observations. 52 stars in total were observed using CSHELL in 2016A and 2016B. The final spectra were normalized and we achieved S/N ratios above 100 for most targets with individual S/N ratios listed in Table \ref{table::obslog}. 

\begin{figure}[tp!]
	\centering 
 	\includegraphics[trim=0cm 0cm 0cm 0cm, scale=.45]{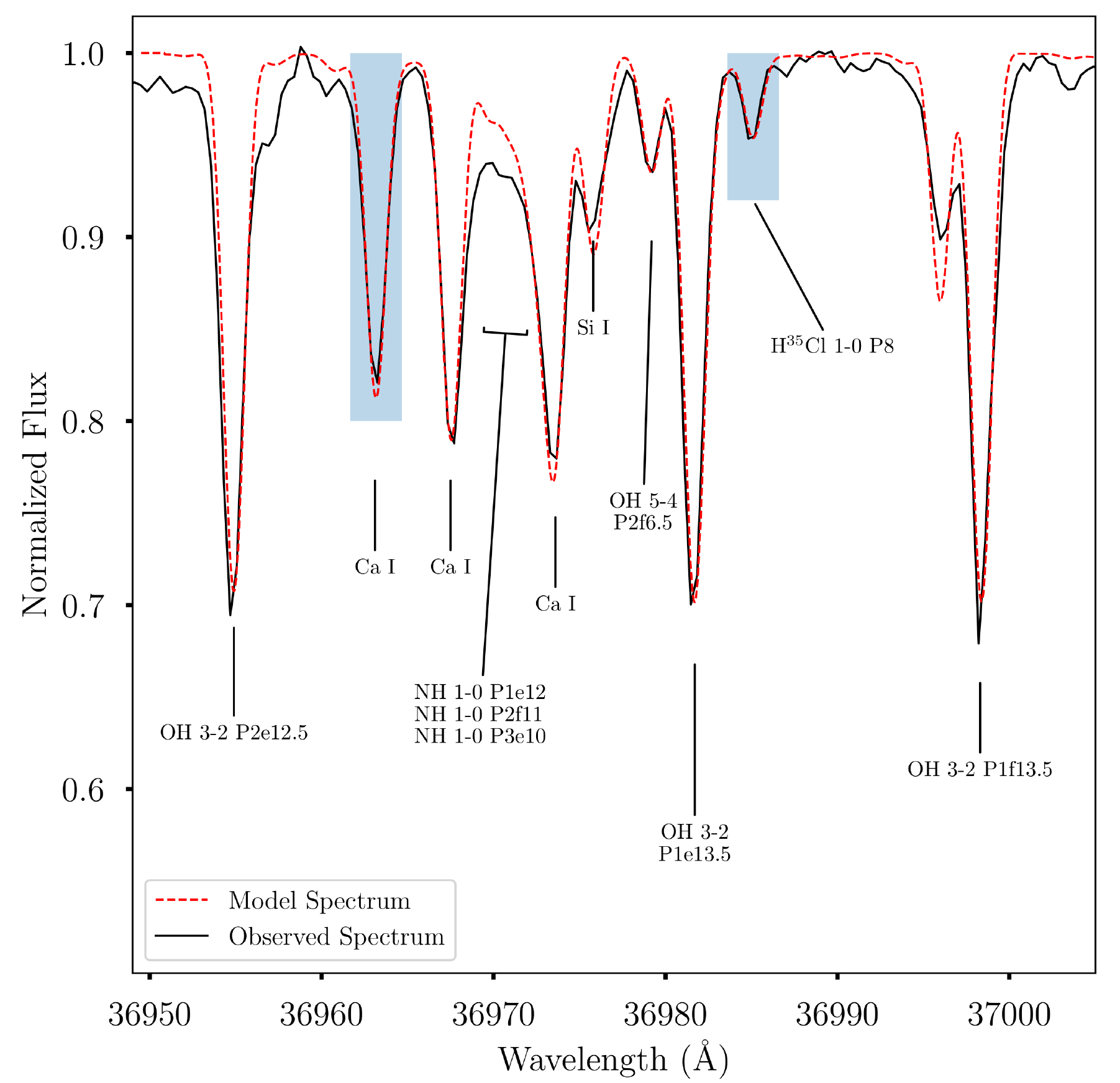}
	\caption{An example CSHELL spectrum of the star HD 123303. The black line is the observation while the red line is a model spectrum. Important atomic and molecular lines are identified. The blue shaded region represents the two lines used for abundance analysis and the wavelengths where the $\chi^{2}$ minimization was performed (see section \ref{subsec:ir_abun}). \label{fig:ir_spec} }
	\end{figure}

\subsection{Optical Spectra Observations and Data Reduction}

The optical echelle spectra were obtained from two different facilities. First, we used ARCES \citep{wang03} on the 3.5m Astrophysical Research Consortium (ARC) telescope at the Apache Point Observatory (APO). ARCES covers a spectral range between 3600 $\mbox{\AA}$ - 10000 $\mbox{\AA}$ with an approximate resolution of R $\sim$ 30,000. ARCES data were collected primarily for fainter objects ( V $\gtrsim$ 7 mag). The brighter targets in our sample were observed with TOU on the Dharma 50" Telescope \citep{ge16}. TOU covers a similar spectra range between 3800 $\mbox{\AA}$ - 9000 $\mbox{\AA}$ with a higher spectral resolution at R $\sim$ 100000. The source of each target's optical spectrum observation is listed in Table \ref{table::obslog}. 

The data reduction for observations taken at both observatories was completed using IRAF\footnote{IRAF is distributed by the National Optical Astronomy Observatory, which is operated by the Association of Universities for Research in Astronomy, Inc., under cooperative agreement with the National Science Foundation.}. The echelle data reduction followed standard steps and we followed the ARCES data reduction guide\footnote{Accessed at \url{https://www.apo.nmsu.edu/arc35m/Instruments/ARCES/}} for data acquired at the ARC 3.5m telescope. During observations, bias, dome flats, and comparison ThAr spectra were observed each night. Blue and red flat field images were observed at both observatories. Cosmic rays were removed from the echelle spectra using the task \texttt{cosmicrays} and median combined bias images were used to subtract the bias level off all other images. Apertures were defined using the flatfield images and were extracted using \texttt{apall}. Next the object images were extracted and inter-order scattered light was removed using the task \texttt{apscatter}. The extracted object spectra were then flatfielded. ThAr reference spectra were used as the wavelength solution and  chebyshev polynomials were used to fit the two dimensional wavelength solution in each image. The typical RMS uncertainty of the wavelength solution fit to the ThAr lamp lines was $\sim$ 0.005 $\mbox{\AA}$ for ARCES data and $\sim$ 0.003 $\mbox{\AA}$ for TOU data. Orders used for abundance analysis were normalize individually using the \texttt{continuum} task. The final signal-to-noise ratios are given in Table \ref{table::obslog} and calculated in a similar way to the IR data although the regions without stellar absorption features were extremely limited. The regions of 7379.79 $\mbox{\AA}$ - 7380.535 $\mbox{\AA}$, 7458.7 $\mbox{\AA}$ - 7458.82 $\mbox{\AA}$, and 7534.6 $\mbox{\AA}$ - 7537.28 $\mbox{\AA}$ were used in the signal-to-noise ratio determinations.

\section{Atmospheric Parameter Determination}
\label{sec::atmo_params}
Typically, atmospheric parameters are derived using Fe I and Fe II excitation and ionization balance. However, the spectral range of our CSHELL spectra is very limited and few Fe II lines exist in the M giant optical spectra. Another difficulty is that strong molecular features in the optical spectra (e.g. TiO) limit the number of unblended absorption features. Also, in M giant stars, low excitation potential Fe I lines are subject to NLTE effects \citep{ruland80,tomkin83,smith90}. We therefore derived the atmospheric parameters from a combination of SED fits to photometry from the literature, luminosities of the stars, and high excitation Fe I lines in our optical spectra. 

\subsection{Effective Temperature}
\label{subsec::teff}
The effective temperatures were adopted from the spectral energy distribution (SED) fits performed by \citet{mcdonald12}. The SED fits used model atmospheres with photometry ranging from the near UV to 100 $\mu$m. Effective temperatures from \citet{mcdonald12} were tested in \citet{maas16} and were found compatible with measurements from spectroscopy. We further tested the \citet{mcdonald12} parameters by comparing the effective temperatures to different color-$T_{\mathrm{eff}}$ relationships.  

\begin{figure}[tp!]
	\centering 
 	\includegraphics[trim=0cm 0cm 0cm 0cm, scale=.4]{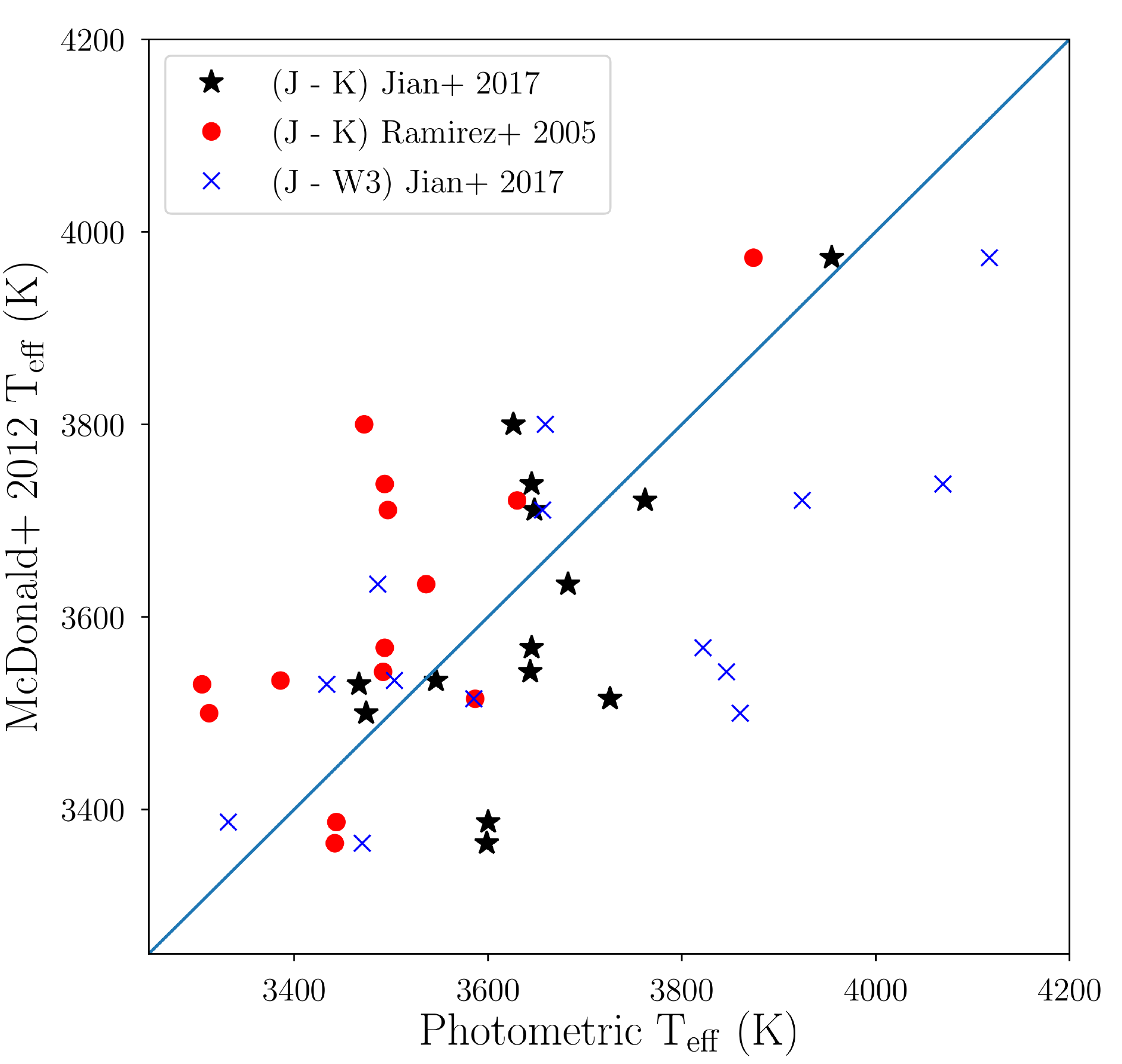}
	\caption{Temperature comparisons for a selection of M giants described in subsection \ref{subsec::teff}. The photometric effective temperature on the x axis comes from three sources: black stars represent (J-K) from \citet{jian17}, red circles from \citet{ramirez05}, and blue 'x's for (J - W3) from  \citet{jian17}. Y axis temperatures are from \citet{mcdonald12}. The blue line represents a line with a slope of 1.  \label{fig:temp_comp} }
	\end{figure}

We derived effective temperatures by extrapolating the J - K color-$T_{\mathrm{eff}}$ from \citet{ramirez05} and the J - K$_{s}$ and J - W3 color-$T_{\mathrm{eff}}$ from \citet{jian17} beyond the initial temperature range of their initial data. The lower limit of the $T_{\mathrm{eff}}$ range explored is 4050 K for \citet{ramirez05} and 3650 K for \citet{jian17}. V band photometry was adopted from the Tycho-2 survey \citep{esa97}, J and K$_{s}$ magnitudes from DIRBE \citep{hauser98,smith04} or 2MASS \citep{skrutskie06} if unavailable from DIRBE, and finally W3 photometry from WISE \citep{wright10}. The  K$_{s}$ photometry was transformed to the Johnson K filter using the prescription from \citet{johnson05}. We then removed any stars with significant dust attenuation (A$_{v}$ $>$ 0.3) estimated by \citet{schlafly11}, significant infrared photometric errors (kept stars with $\sigma_{J}$, $\sigma_{K}$ $<$ 0.3 mag), and removed the coolest stars from the sample ($T_{\mathrm{eff}}$ $<$ 3300 K) where the extrapolation may be the least reliable. The results of the temperature comparison are found in Fig. \ref{fig:temp_comp}. 

The \citet{ramirez05} temperature relations are systematically offset from the adopted temperatures while the \citet{jian17} results appear consistent with our adopted temperatures from SED fits. The average difference and standard deviation for the two color-$T_{\mathrm{eff}}$ relations from \citet{jian17} are 35 K $\pm$ 117 K for J - K$_{s}$ and 89 K $\pm$ 173 K for J - W3. We therefore adopt the \citet{mcdonald12} temperatures with uncertainties of $\pm$ 100 K. 

\subsection{Surface Gravity}

Next, we calculated the surface gravity for each star in our sample using Eq. \ref{eq::logg}:

\begin{equation}
\label{eq::logg}
\mathrm{Log}(g/g_{\odot}) = \mathrm{Log}(M/M_{\odot}) - \mathrm{Log}(L/L_{\odot}) + \mathrm{4Log}(T/T_{\odot})
\end{equation}

\noindent In equation \ref{eq::logg}, M is the stellar mass, L is the luminosity, and T is the effective temperature. The luminosity of each star was derived using the K magnitude from DIRBE or 2MASS \citep{hauser98,smith04} (transformed to Johnson K), distances using Gaia DR2 parallaxes \citep{gaia18} from \citet{bailer18}, and K band bolemetric corrections from \citet{bessell98}. Due to the spacing in the grid of bolometric corrections for the M giant stars, we adopt an uncertainty of $\Delta$BC$_{K}$ = $\pm$ 0.2 when calculating the luminosity uncertainty. For stars with no available Gaia DR2 parallax, Hipparcos parallax measurements \citep{esa97} were used instead to calculate the luminosity. 

\begin{figure}[h!]
	\centering 
 	\includegraphics[trim=0cm 0cm 0cm 0cm, scale=.44]{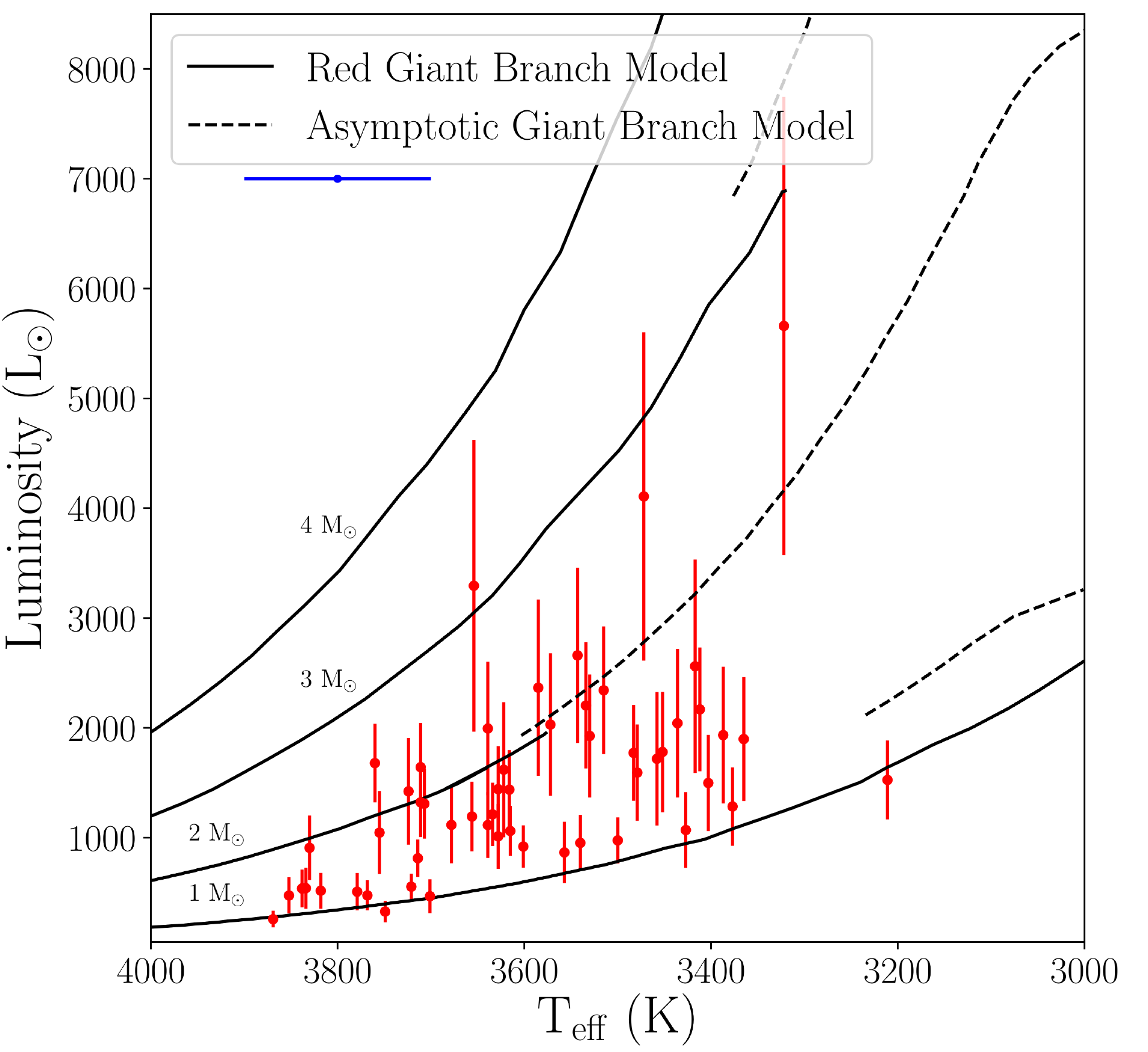}
	\caption{A HR diagram for the M giant sample. Black lines are red giant branch stellar evolution tracks and dashed lines are asymptotic giant branch stellar evolution tracks, both from \citet{bertelli08,bertelli09}. A representative effective temperature error represented by a blue bar. \label{fig:hr_diag} }
	\end{figure}

With known luminosities and effective temperatures we compared our results to stellar evolutionary tracks from \citet{bertelli08,bertelli09}. Specifically, we assumed our M giants have metallicities consistent with the solar Z and used the tracks with Y = 0.30 and Z = 0.017. The effective temperature and luminosities for each star were compared to stellar evolution tracks at various masses and the model that most closely matched the two parameters was adopted as the stellar mass of the star. The models between 0.8 - 2.2 M$_{\odot}$ varied in steps of 0.1 M$_{\odot}$ and 2.5 - 4  M$_{\odot}$ varied in steps of 0.5  M$_{\odot}$. We tested the uncertainty on the mass calculations by conducting a Monte Carlos simulation for a typical star in our sample, HD 164448 ($T_{\mathrm{eff}}$ = 3634, L = 1210 L$_{\odot}$, and mass = 1.7 M$_{\odot}$). We re-computed the effective temperature and luminosity using a Gaussian number with the standard deviation set to the uncertainties, recalculated the mass, and repeated 5,000 iterations. The standard deviation on the distribution was 0.32 M$_{\odot}$. Since the stars with larger masses have have sparser mass coverage, we estimated the uncertainty on each star to $\delta$M = $\pm$0.5 M$_{\odot}$ when calculating the log(g). Fig. \ref{fig:hr_diag} comparison of the luminosity and effective temperature to some stellar evolutionary tracks. We find most stars in our sample range between 1 M$_{\odot}$ and 2 M$_{\odot}$. Finally log(g) values are calculated with Eq. \ref{eq::logg} with uncertainties from each intrinsic parameter propagated and the average log(g) uncertainty is 0.20 $\pm$ 0.03 for the sample.  

\subsection{[Fe/H] and Microturbulence}
The [Fe/H] abundance for each star was measured using Fe I lines in the optical echelle spectra between 7400 $\mbox{\AA}$ - 7580 $\mbox{\AA}$ and from 8600 $\mbox{\AA}$ to 8700 $\mbox{\AA}$. The chosen spectral regions have both minimal telluric contamination and no strong TiO molecular bands. Fe I lines used between 7400 $\mbox{\AA}$ - 7580 $\mbox{\AA}$ have previously been used to determine Fe I abundances in M giant stars \citep{smith90,chou07}. We chose atomic line information (e.g. excitation potential) from the Kurucz database\footnote{http://kurucz.harvard.edu/atoms.html} and lines from \citet{chou07}. 

\begin{deluxetable}{ c c c c }
\tablewidth{0pt} 
\tabletypesize{\footnotesize}
\tablecaption{Atomic Ti I and Fe I lines \label{table::lines}} 
 \tablehead{\colhead{Species} & \colhead{Wavelength} & \colhead{$\xi$} & \colhead{log(gf)}\\
 \colhead{} & \colhead{($\mbox{\AA}$)} & \colhead{(eV)}& \colhead{} }
\startdata
Ti I & 7402.675	&	3.179&	-0.973\\
Ti I & 7416.99	&   1.067&	-3.345\\
Ti I & 7423.183	&	1.443&	-2.612\\
Ti I & 7432.67	&   1.460&	-2.751\\
Ti I & 7456.584 &	0.818&	-3.468\\
Ti I & 7469.938	&	0.836&	-3.268\\
Ti I & 7491.362	&	0.826&	-3.553\\
Fe I\tablenotemark{a} & 7440.911	&	4.913 &	-0.572 \\
Fe I & 7443.022	&4.186	&-1.75    \\
Fe I\tablenotemark{a} & 7445.749 &	4.256 & -0.103	\\
Fe I & 7447.384	&4.958	&-1.011   \\
Fe I & 7461.519	&2.559	&-3.53    \\
Fe I\tablenotemark{a} & 7476.375 &	4.795 &	-1.55\\
Fe I & 7498.53	&4.143	&-2.19    \\
Fe I & 7507.265	&4.415	&-0.972   \\
Fe I & 7511.018	&4.178	&0.187    \\
Fe I & 7531.143	&4.371	&-0.67    \\
Fe I & 7547.896	&5.099	&-1.147   \\
Fe I & 7568.898	&4.283	&-0.822   \\
Fe I & 8611.803	&2.845	&-1.86    \\
Fe I & 8674.746	&2.831	&-1.81    \\
Fe I & 8698.706	&2.990	&-3.517   \\
Fe I & 8699.455	&4.955	&-0.469   \\
\enddata
\tablenotetext{a}{Line only measured in TOU Optical Spectra}
\end{deluxetable}

To test our linelist choices, we measured equivalent widths in the Sun \citep{wallace11} and Arcturus \citep{hinkle00} using visible solar atlases for all detectable Fe I lines in the spectral region. The log(gf) values were adjusted for some Fe I lines to match the solar and Arcturus abundance. To avoid using spectral lines subjected to large systematic effects with temperature or luminosity, any line that could not simultaneously reproduce the abundance in Arcturus and in the Sun was removed from the linelist. We derived an abundance of A(Fe I) = 7.51 $\pm$ 0.05 (st.dev.) for the Sun and A(Fe I) = 6.97 $\pm$ 0.05 (st.dev.) for Arcturus with the linelist in Table \ref{table::lines}, consistent with the values from \citet{asplund09} and \citet{ramirez11} respectively. 

 The Fe I lines were examined in the R $\sim$ 100,000 spectra and R $\sim$ 30,000 ARCES spectra and lines with significant blending were removed from the linelist. The final set of lines with astrophysical log(gf) values are listed in Table \ref{table::lines}. The lines beyond 7580 $\mbox{\AA}$ were not available in the TOU spectra while three lines heavily blended in the ARCES spectra were measurable in the higher resolution TOU spectra. Those lines are indicated in Table \ref{table::lines}. 

\subsubsection{Determining [Fe/H] and Microturbulence}
Equivalent width measurements were then performed for all stars using \texttt{splot} task within IRAF. Iron abundances for each line were calculated using the \texttt{abfind} routine with MOOG spectral synthesis software (\citealt{sneden73}; V. 2017) and MARCS model atmospheres \citep{gustafsson08}. Models were initially set to the [Fe/H] = 0 and a microturbulence value of 2.0 km s$^{-1}$. The best microturbulence value was calculated by conducting a grid search in steps of 0.01 km s$^{-1}$ and determining which microturbulence value minimized the slope of the A(Fe I) abundance and reduced equivalent width  (log(EW/$\lambda$)) for all Fe I lines measured. The process was repeated with a new atmospheric model, created with the average A(Fe I) abundance and microturbulence from the last iteration. The iterations were stopped and the final values chosen, when the A(Fe I) abundance did not changed by more than 0.01 dex. Typically only two to three iterations were necessary. The final set of atmospheric parameters is given in Table \ref{table::mgiant_results}.

The standard deviation calculated from the abundances derived for individual Fe I lines determined was adopted as the statistical [Fe/H] uncertainty. Additionally, uncertainties in the effective temperature and surface gravity were calculated for five sample stars, then applied the average uncertainty to the rest of the Fe abundances. For the five stars, models at $\pm$ 100 K and $\pm$ 0.20 in log(g) were created individually, the new A(Fe I) computed, and the difference adopted as the uncertainty. The atmospheric parameter uncertainties were added in quadrature with the uncertainty on the fit. The final average $\delta$A(Fe I) is $\pm$ 0.16 dex. The microturbulence uncertainty was calculated using the 1$\sigma$ uncertainty on the best fitting linear line to the A(Fe I) abundance and reduced equivalent width (log(EW/$\lambda$)), obtained from the covariance matrix. The uncertainty in the microturbulence is the difference between the grid search result and the microturbulence corresponding to the 1$\sigma$ slope. The average difference is $< \Delta \chi >$ = 0.15 $\pm$ 0.03 (s.t.dev) for the five stars. We used the average 2$\sigma$ value of 0.30 kms$^{-1}$ as the uncertainty estimate for the entire sample to ensure no stars were beyond this limit in our sample. 

\begin{deluxetable*}{cccccccccc}
\tablewidth{0pt} 
\tabletypesize{\footnotesize}
\tablecaption{M Giant Stellar Parameters and Abundances \label{table::mgiant_results}} 
 \tablehead{\colhead{HD} & \colhead{$T_{\mathrm{eff}}$} & \colhead{log(g)} & \colhead{A(Fe I)} & \colhead{[Fe/H]} & \colhead{$\xi$} & \colhead{A(O)} & \colhead{A($^{35}$Cl)} & \colhead{A(Ca)} & \colhead{A(Ti)}    \\
 \colhead{Number} & \colhead{(K)} & \colhead{} & \colhead{} & \colhead{}& \colhead{(km s$^{-1}$)}  & \colhead{}& \colhead{} & \colhead{}& \colhead{}}
\startdata
5111	&	3636	&	0.91	&	7.51	&	0.01	&	2.13	&	\nodata{}	&	4.98	&	6.29	&	4.93	\\
15594	&	3436	&	0.43	&	7.40	&	-0.10	&	1.71	&	\nodata{}	&	4.71	&	6.01	&	4.67	\\
19258	&	3601	&	0.79	&	7.29	&	-0.21	&	1.89	&	\nodata{}	&	4.73	&	6.18	&	4.63	\\
25000	&	3479	&	0.56	&	7.52	&	0.02	&	1.91	&	\nodata{}	&	4.88	&	6.24	&	4.87	\\
30634	&	3714	&	0.94	&	7.26	&	-0.24	&	2.01	&	8.46	&	4.43	&	5.85	&	4.63	\\
31072	&	3779	&	1.11	&	7.46	&	-0.04	&	2.13	&	8.64	&	5.15	&	6.22	&	4.78	\\
35067	&	3701	&	1.00	&	7.50	&	0.00	&	2.00	&	8.58	&	4.95	&	6.11	&	4.79	\\
84181	&	3622	&	0.71	&	7.35	&	-0.15	&	1.87	&	\nodata{}	&	5.09	&	6.22	&	4.74	\\
99592	&	3365	&	0.37	&	7.28	&	-0.22	&	1.82	&	\nodata{}	&	4.56	&	5.62	&	4.65	\\
102159	&	3387	&	0.40	&	7.44	&	-0.06	&	2.22	&	\nodata{}	&	4.69	&	6.01	&	4.69	\\
\enddata
\tablecomments{(This table is available in its entirety in machine-readable form.)}
\end{deluxetable*}

\section{Abundance Analysis}
\label{sec:abun_measure}
\subsection{Abundance Derivations}
\label{subsec:ir_abun}
Abundances were derived using MOOG spectral synthesis software (\citealt{sneden73}; v. 2017) with MARCS 1-D atmospheric models \citep{gustafsson08}. We adopted the linelist from \citet{maas16} to determine abundances from the IR spectra. In the previous studies, CNO abundances were derived using CH, OH, and NH molecular features at 3.7 $\mu$m. However, the very weak CH features were too significantly blended to discern in the lower resolution CSHELL spectra. The molecular equilibrium for molecules containing CNO depends on the abundances of all three species. Without independent methods to determine all CNO abundances we focus on determining only the Cl and Ca abundances from our IR spectra. Ca provides an accessible $\alpha$ element for comparison with Cl and abundance measurements in disk stars using Ca I features at 3.7 $\mu$m are consistent with abundances from optical lines \citep{maas16}.

We determined which abundance minimized $\chi^{2}$ for the multiple sets of synthetic spectra and observed IR spectrum of each star. The $\chi^{2}$ minimization was done over the shaded regions in Fig. \ref{fig:ir_spec} to find the abundances of Ca and Cl. We avoided the two blended Ca features since we are not deriving CNO abundances. A grid search was performed using the synthetic spectra created over a range of Cl and Ca abundances. The best fitting abundances for the two regions are given in Table \ref{table::mgiant_results}. We use \citet{asplund09} as the solar reference for all elements except for Cl in which we use the meteoric value from \citet{lodders09}, consistent with the methodology of \citet{maas16}.

\subsection{Oxygen Abundance Derivations}

Oxygen abundances are necessary to meaningfully compare our results to PN and H II regions. While OH lines are present in the L band spectra, determining abundances from them without carbon abundances is difficult due to the CNO molecular equilibrium. We therefore use the forbidden oxygen line at 6300 $\mbox{\AA}$ to determine oxygen abundances in our stars. The 6300 $\mbox{\AA}$ region is on the edge of a TiO band centered at $\sim$ 6150 $\mbox{\AA}$. A linelist was created using the [O I] line, nearby Fe I lines, and TiO lines from the $\delta$ transition taken from \citet{plez98}. Due to the number of of TiO lines, only lines with log(gf) values above --0.5 and excitation potentials lower than 1 eV were selected. Additional lines contributed little to the resulting synthetic spectra while increasing computation time. Solar Ti isotope fractions were assumed and adopted from \citet{asplund09}. Synthetic spectra were created using MOOG (\citealt{sneden73}; v. 2017) and MARCS atmospheric models \citep{gustafsson08}. The linelist was tested on Arcturus and reproduces the oxygen abundance and Fe abundance from \citet{ramirez11}.

\begin{figure}[tp!]
	\centering
	\includegraphics[trim=0cm 0cm 0cm 0cm, scale=.35]{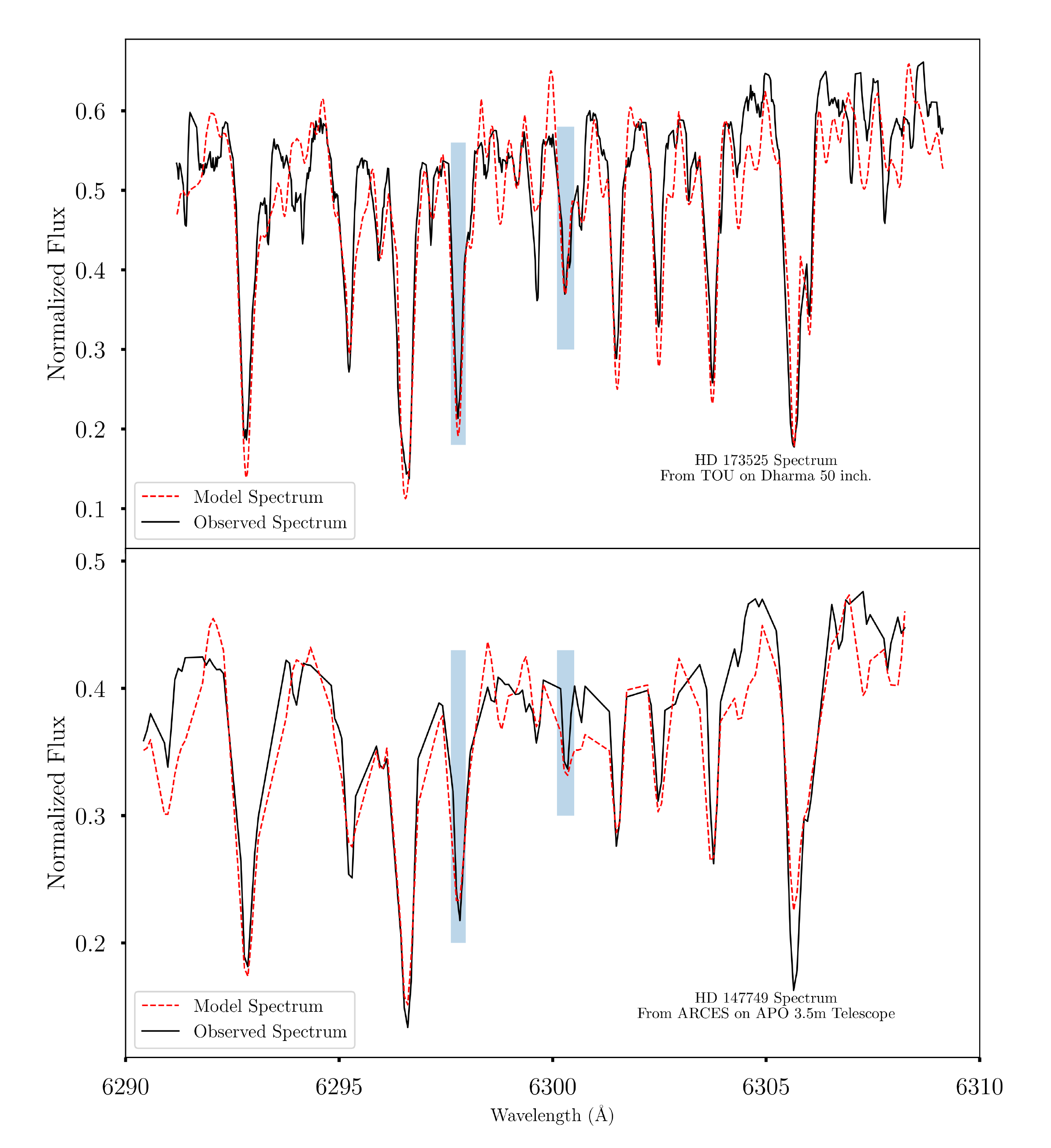}
	\caption{Both panels show synthetic spectra fits to observed data. The blue boxes show the features chosen for the $\chi^{2}$ minimization. Top Panel: HD 173525 spectrum fit. The star was observed using TOU on the Dharma 50" telescope. Bottom Panel: HD 147749 spectrum fit. The star was observed using ARCES on the APO 3.5m telescope. \label{fig:hcl_optical_oxy_spectra}  }
	\end{figure}

The stellar continuum was difficult to determine due to the TiO molecular lines blanketing the spectral region. First, we used the \texttt{continuum} task in IRAF and normalized to the apparent continuum in the [O I] spectral region. A synthetic spectrum with only the TiO component was created and fit with a spline function. The initially divided spectra in our sample were then divided by the TiO spline fit to re-introduce the TiO component across the wavelength range of interest. Best fit spectra were determined by performing a $\chi^{2}$ minimization over the [O I] line region and a nearby Fe I line to constrain the fit to the continuum from TiO. A grid of synthetic spectra that varied in oxygen abundances in 0.01 steps were created and fit; re-calculating the TiO contribution to the continuum at every step. The $\chi$ values were also weighted by the TiO continuum contribution to ensure the minimum $\chi^{2}$ value was not related to the decreasing flux with higher O abundances, and stronger TiO features. Examples of the best fitting spectra for HD 147749 and HD 173525 are shown in Fig. \ref{fig:hcl_optical_oxy_spectra}. Abundance uncertainties from the atmospheric parameters were calculated using the same methodology as subsection \ref{subsec:uncer_atmoparams}.

 	\begin{figure}[tp!]
	\centering
	\includegraphics[trim=0cm 0cm 0cm 0cm, scale=.33]{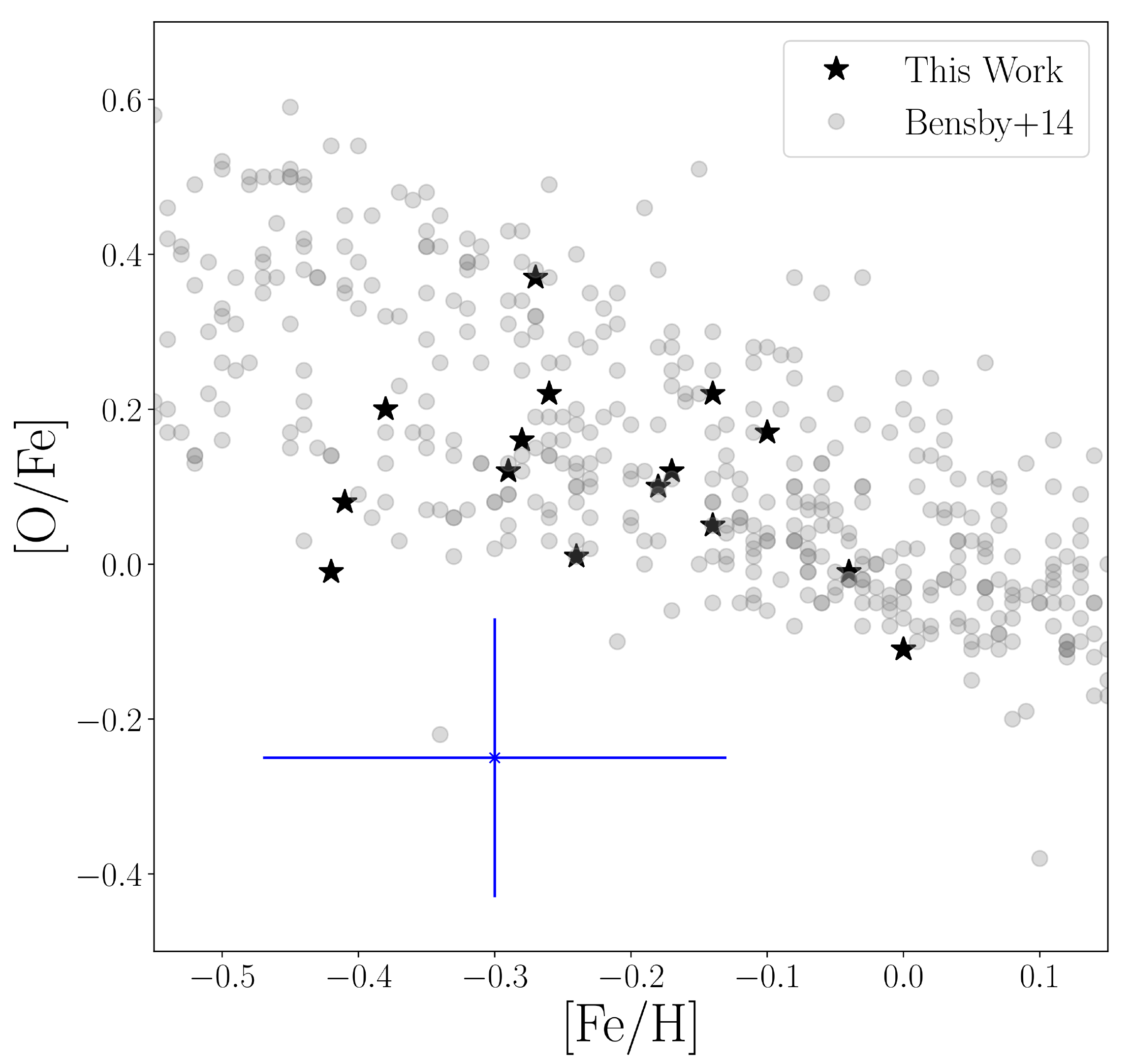}
	\caption{[O/Fe] ratios vs. [Fe/H]. Black stars are abundance ratios from this work and grey circles are from \citet{bensby14}. The blue point give representative error bars on the Cl and O abundances.
 \label{fig:mgiant_oxy_bensby}  }
	\end{figure}
	
Additionally, telluric lines are present near the spectral region containing the [O I] feature. We did not perform a telluric correction for the optical spectra and although we do not see any anomalously strong [O I] features or blends with un-identified lines, contamination still may be present. Only a subsample of our observed stars with $T_{\mathrm{eff}}$ $>$ 3700 K have O abundances measured. Below this temperature, the TiO features begin dominating all other features and systematic uncertainties from modeling TiO continuum, specifically the linelist and atmospheric parameters, become significant. We therefore measured O abundances in the 15 warmest stars from the original sample. The oxygen abundances are listed in Table \ref{table::mgiant_results} with average uncertainties listed in Table \ref{table::maas_hcl_uncertainties}. We also compare our derived oxygen abundances to [O/Fe] ratios in FGK stars from \citet{bensby14} in Fig. \ref{fig:mgiant_oxy_bensby}. We found our [O/Fe] ratios agree with ratios from \citet{bensby14}, providing evidence our methodology provides sufficiently accurate results to compare to nebular abundances. 

\subsection{Uncertainty Analysis} 

\subsubsection{Uncertainty from Fit}
The noise in the final, reduced, 1D spectra come from multiple sources including the original counts, division with a telluric spectrum, and division by a flat-field spectrum. We devised multiple similar methodologies to account for this source of uncertainty when fitting synthetic spectra to the data. 

The uncertainty in the Cl and Ca spectral synthesis fit was estimated through a Monte Carlo simulation. A grid of synthetic spectra with an individual element abundance change in steps of 0.01 dex for each spectrum in the grid, was created for Ca and $^{35}$Cl. A new spectrum was created by generating Gaussian random numbers for each pixel flux value with the mean value the original measured flux and a standard deviation equal to the signal-to-noise ratio of the data. The best fitting elemental abundances were calculated for each adjusted stellar spectrum. The process was repeated for 100,000 iterations and the 16$\%$ and 84$\%$ values of the posterior distribution of the Monte Carlo simulation were adopted as the uncertainty on the fit. For Ti and Fe, abundances were individually derived for multiple lines with equivalent width measurements. The average of the lines was taken as the stellar abundance and the standard deviation as the uncertainty due to the fit.

\subsubsection{Uncertainty from Atmospheric Parameters}
\label{subsec:uncer_atmoparams}
The uncertainty on each abundance measurement due to noise in the spectra and uncertainty in the atmospheric parameters, assumed to be independent, was added in quadrature. The uncertainty for each abundance due to uncertainty on the adopted atmospheric parameters was calculated by using new atmospheric models. We varied one atmospheric parameter by 1$\sigma$ (e.g. $T_{\mathrm{eff}}$) while the others were held constant to the adopted values, created a new atmospheric model, then new abundances were derived. The best fitting abundances were found by minimizing the $\chi^{2}$ between the model fit and the synthetic spectra, the same procedure described in subsection \ref{subsec:ir_abun}.  The 1$\sigma$ uncertainties from the atmospheric parameters is defined as the difference between the abundance derived and the atmospheric models with adjusted atmospheric parameters and are listed in Table \ref{table::maas_hcl_uncertainties}. 

\begin{deluxetable}{ c c c c c}
\tablewidth{0pt} 
\tabletypesize{\footnotesize}
\tablecaption{M Giant Abundance Uncertainty Averages \label{table::maas_hcl_uncertainties}} 
 \tablehead{\colhead{Uncertainty} &  \colhead{$\delta$A(O)} & \colhead{$\delta$A($^{35}$Cl)} & \colhead{$\delta$A(Ca)} & \colhead{$\delta$A(Ti)} \\ \colhead{Source} & \colhead{(dex)} & \colhead{(dex)} & \colhead{(dex)}& \colhead{(dex)}}
\startdata
$T_{\mathrm{eff}}$  &  0.04 & 0.20 & 0.04 & 0.09 \\
log(g) & 0.04 & 0.01 & 0.03 & 0.04\\
$\mathrm{[Fe/H]}$ & 0.07 & 0.06 & 0.17 & 0.09\\
$\xi$ & 0.15 & 0.00 & 0.07 & 0.10\\
Fit & 0.03 & 0.05 & 0.02 & 0.10\\
Tot. Uncertainty & 0.18 & 0.21 & 0.19 & 0.19 \\
\enddata
\end{deluxetable}

\subsubsection{Uncertainty with Effective Temperature}
\label{subsec:mgiant_systematic_atmo_params}
No NLTE calculations are available for the L band HCl transitions. Additionally, blends with weak, un-identified molecular features may impact our abundances. In both scenarios, the systematic uncertainty will likely change as a function of the effective temperature. \citet{maas16} found no relationship between $^{35}$Cl, C, O, N, Si, and Ca with $T_{\mathrm{eff}}$ using abundances from the same L band region. We test those results by examining the relationship between the adopted $T_{\mathrm{eff}}$ and abundance with the our larger sample of 52 M giants.

To check for consistency within our abundance measurements, we examined how the atmospheric parameters correlated each another. While most atmospheric parameters are uncorrelated, there is a relationship between the effective temperature and [Fe/H], where cooler stars are more metal rich compared to the warmer stars in the sample. The average abundance for stars with effective temperatures cooler than 3500 K is $<$[Fe/H]$>$ = --0.09 while the stars with $<$[Fe/H]$>$ = --0.19 for stars warmer than 3500 K. The offset may be due to unknown systematics with the [Fe/H] calculations, such as unknown blends with molecular lines, from problems with the atmospheric models, or due to a stellar evolution effect. Higher metallicity stars evolve with relatively cooler effective temperatures on the red giant branch (e.g. \citealt{bertelli08}). Therefore, we may expect a metallicity trend with effective temperature in our sample since the cooler stars may be more likely to be metal-rich when selected by effective temperature and magnitude.   

\begin{figure*}[tp!]
	\centering 
 	\includegraphics[trim=0cm 0cm 0cm 0cm, scale=.4]{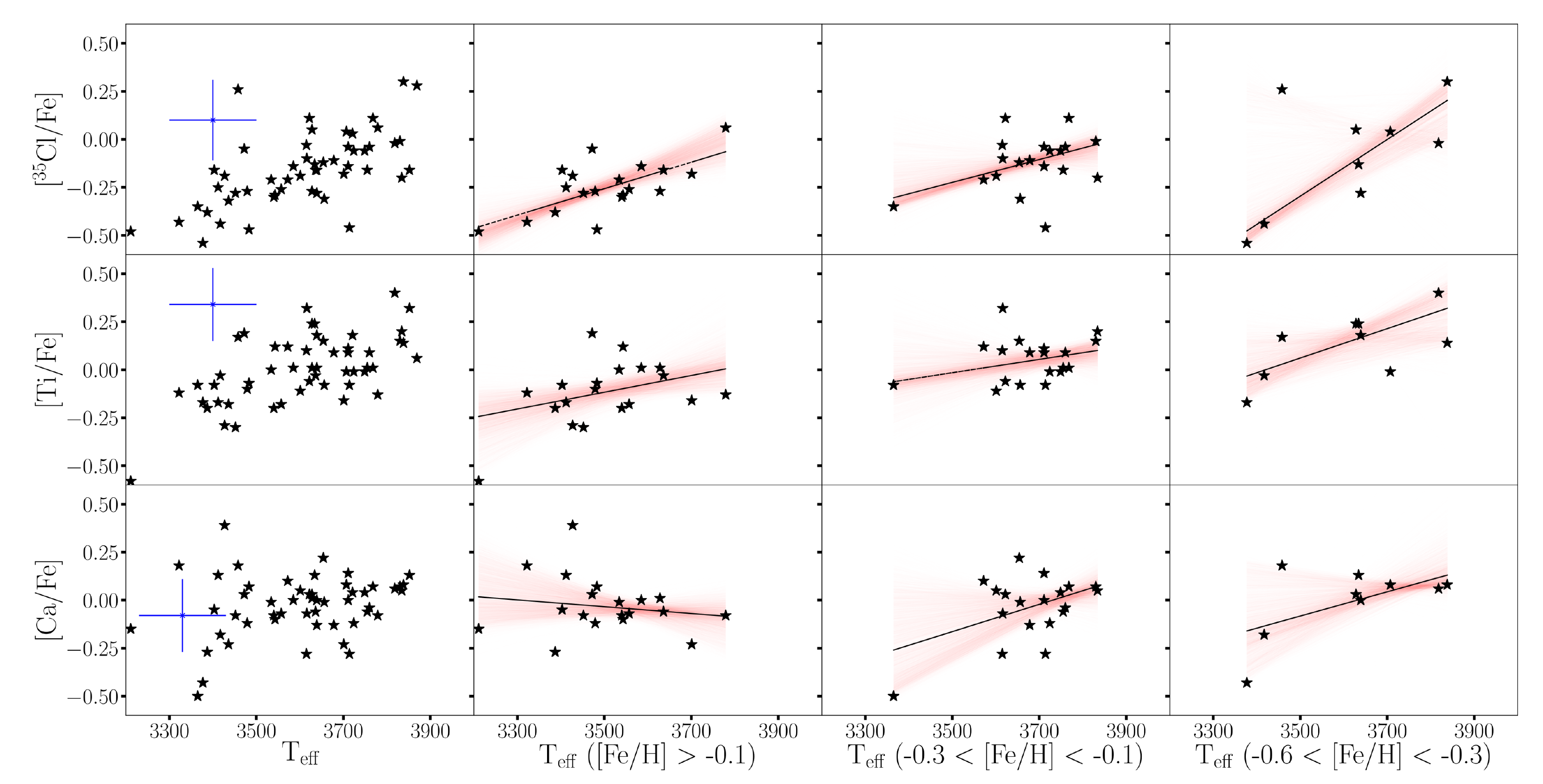}
	\caption{Plots of [$^{35}$Cl/Fe] (top), [Ti/Fe] (middle), and [Ca/Fe] (bottom) versus $T_{\mathrm{eff}}$. Blue crosses are representative error bars. Left most plot shows the entire sample of stars. Next plots are binned in [Fe/H] $>$ --0.1, --0.3 $<$ [Fe/H] $<$ --0.1, and --0.6 $<$ [Fe/H] $<$ --0.3 respectively. Solid black line is best linear fit to the data and the red lines in each bin represent 3000 bootstrap fits. The slopes for each parameter are given in Table \ref{table::[x/fe]_vs_teff}.  \label{fig:teff_cl_ti_ca_correlation}}
	\end{figure*}

One way to determine the cause of [Fe/H] and $T_{\mathrm{eff}}$ relation is to look at the relationship between other elements with effective temperature in the M giant sample. We plot the effective temperature against the [$^{35}$Cl/Fe], [Ca/Fe], and [Ti/Fe] ratios in Fig. \ref{fig:teff_cl_ti_ca_correlation}. We find both [Ti/Fe] and [$^{35}$Cl/Fe] increase with increasing effective temperature. To further understand this relationship, we split the sample into three metallicity bins and do a bootstrap estimate to determine the relationship between abundance and effective temperature. The bin sizes were chosen to approximately match the region before the [Ti/Fe] slope increases in \citet{bensby14}. Bin sizes were also chosen to cover the necessary $T_{\mathrm{eff}}$ range and finally to probe roughly different regimes in the MARCS [Fe/H] grid spacing (--0.5, --0.25, and solar metallicity). We fit each bin with a linear line and the bootstrap was performed with 10,000 iterations. The linear fit was performed with a Levenburg-Marquardt algorithm and soft l1 loss smoothing approximation to remove outliers. The resulting slopes and their uncertainties are found in Table \ref{table::[x/fe]_vs_teff}. We find the [Ti/Fe] and [Ca/Fe] ratios, at similar [Fe/H] ratios, are approximately constant over the $T_{\mathrm{eff}}$, as shown in Fig. \ref{fig:teff_cl_ti_ca_correlation}. When only the stars with $T_{\mathrm{eff}}$ $>$ 3500 K are fit, the slopes become statistically consistent with zero.

Only the high metallicity [$^{35}$Cl/Fe] bin has a significant non-zero slope. However, the $^{35}$Cl abundance increases with effective temperature, which is not consistent with unknown molecular blends becoming stronger as effective temperature decreases, artificially enhancing the H$^{35}$Cl line strength. The warmest star in the bin has a large [$^{35}$Cl/Fe] ratio and has a high leverage on the fit, even when the coolest stars are removed. For the Fe~I lines, if an unidentified molecular feature was blended with either the Ti~I features, Fe~I features, or the H$^{35}$Cl molecular feature, a relationship with effective temperature should be especially present in the highest metallicity bin, in which no clear trend is present in Fig. \ref{fig:teff_cl_ti_ca_correlation}. 

\begin{figure*}[tp!]
	\centering 
 	\includegraphics[trim=0cm 0cm 0cm 0cm, scale=.25]{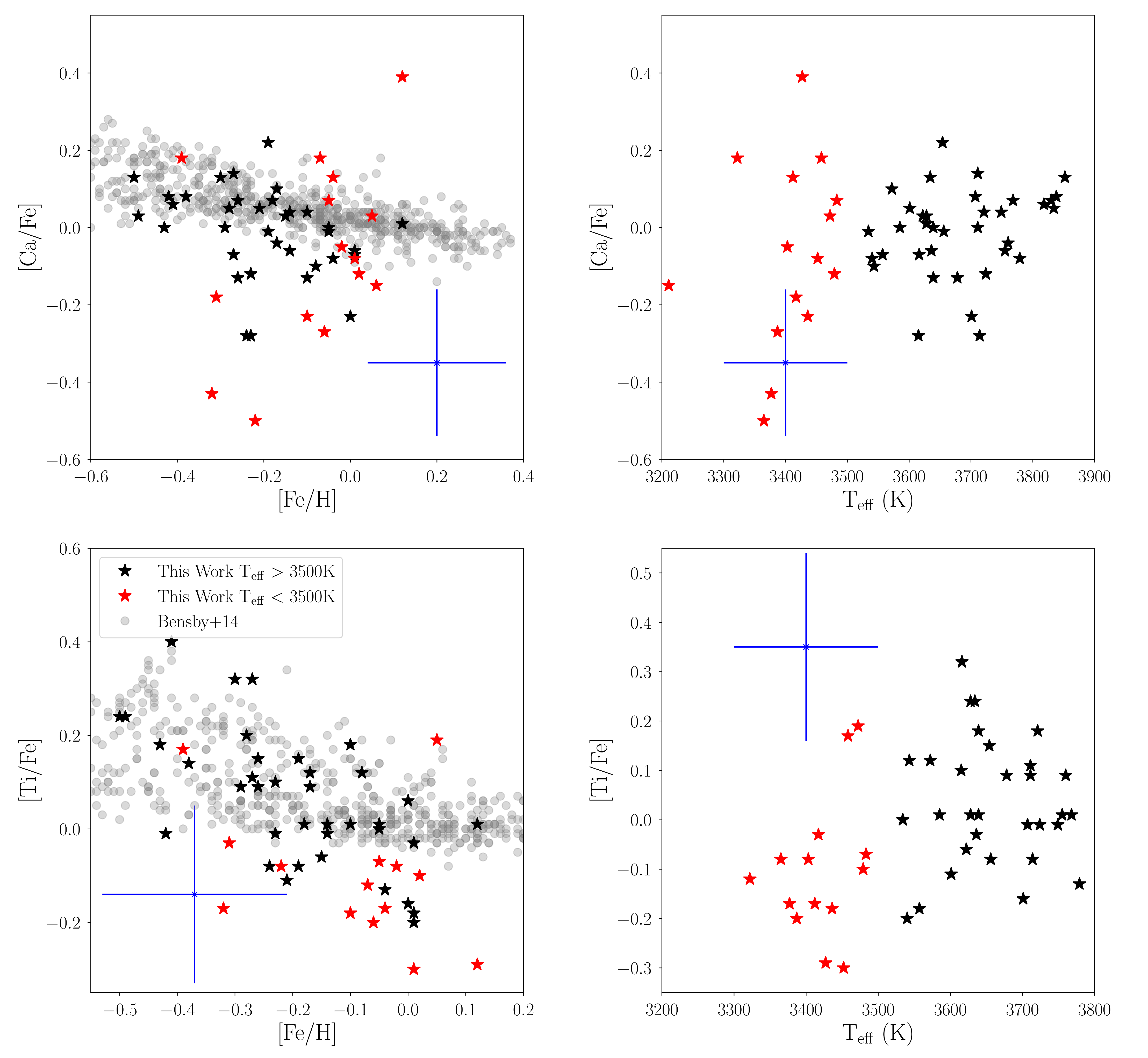}
	\caption{Left: [Ca/Fe] and [Ti/Fe] vs. [Fe/H] ratios. The stars are from the M giant sample, red stars have $T_{\mathrm{eff}}$ $<$ 3500 K while black stars have $T_{\mathrm{eff}}$ $>$ 3500 K. The gray circles are from \citet{bensby14}. Right: The [Ca/Fe] and [Ti/Fe] ratios vs. effective temperature. In all plots, a blue point is shown with representative error bars.  \label{fig:ti_ca_bensby} }
	\end{figure*}

$^{35}$Cl, Ti, and Ca have different average abundance ratios to Fe in each bin. Since Ti and Ca surface abundances should not be affected by stellar evolutionary effects, we expect the abundance ratios of the M giants to be similar to FGK stars in the disk. We can compare the [Ti/Fe] and [Ca/Fe] ratios in our sample with \citet{bensby14} abundances from FGK disk stars, shown in Fig \ref{fig:ti_ca_bensby}. The left panels compare the abundances while the right panels show the [X/Fe] ratios versus effective temperature for the M giant sample from this work.

We find overall agreement between [Ti/Fe] and [Ca/Fe] between both samples, although due to the large uncertainties we cannot distinguish if stars are consistent with thin or thick disk abundance ratios. However, the coolest stars in our sample are systematically lower in the [Ti/Fe] sample. We also calculate the average [Ti/Fe] in the different metallicity bins for our sample and also \citet{bensby14}, with the results shown in Table \ref{table::average_ti}. The averages in Table \ref{table::average_ti} demonstrate that the coolest stars, below 3500 K, are systematically offset and removing them from the sample brings the abundances more in line with literature values. M giant [Ca/Fe] ratios appear systematically lower compared to the abundances of \citet{bensby14}. Only one Ca line is used and difficulties defining the continuum, or problems with the higher excitation energy lines in cooler models, may introduce the offset with \citet{bensby14} results. We note the scatter in [Ca/Fe] ratios is larger for the cooler stars, demonstrated in both panels showing [Ca/Fe] ratios in Fig. \ref{fig:ti_ca_bensby}.

\begin{deluxetable}{ l l c c }
\tablewidth{0pt} 
\tabletypesize{\footnotesize}
\tablecaption{Slopes from Linear fits to [X/Fe] vs $T_{\mathrm{eff}}$ for Fig. \ref{fig:teff_cl_ti_ca_correlation}. \label{table::[x/fe]_vs_teff}} 
 \tablehead{ \multicolumn{1}{c }{[X/Fe]} & \multicolumn{1}{c }{[Fe/H]} & \multicolumn{1}{c}{All Stars} & \multicolumn{1}{c}{Stars with $T_{\mathrm{eff}}$ $>$ 3500K} \\ \multicolumn{1}{c}{} & \multicolumn{1}{c}{Bin} & \multicolumn{1}{c}{Slope}  & \multicolumn{1}{c}{Slope}   \\\multicolumn{1}{c}{} & \multicolumn{1}{c}{} & \multicolumn{1}{c}{(dex 100K$^{-1}$)}  & \multicolumn{1}{c}{(dex 100K$^{-1}$)}}
\startdata
$\mathrm{[Cl/Fe]}$ & [--0.6, --0.3] & 0.10 $\pm$ 0.08	 &		0.1 $\pm$ 0.4 \\
$\mathrm{[Cl/Fe]}$ & [--0.3, --0.1] & 0.05 $\pm$ 0.02	&		0.03 $\pm$ 0.04  \\
$\mathrm{[Cl/Fe]}$ & [--0.1, 0.2]  & 0.09 $\pm$ 0.02	&		0.14 $\pm$ 0.04 \\
$\mathrm{[Ti/Fe]}$ & [--0.6, --0.3] & 0.06 $\pm$ 0.04 	&		--0.01 $\pm$ 0.14 \\
$\mathrm{[Ti/Fe]}$ & [--0.3, --0.1] & 0.03 $\pm$ 0.03	&		0.02 $\pm$ 0.05  \\
$\mathrm{[Ti/Fe]}$ & [--0.1, 0.1]  & 0.04 $\pm$ 0.03	&		0.00 $\pm$ 0.05\\
$\mathrm{[Ca/Fe]}$ & [--0.6, --0.3] & 0.06 $\pm$ 0.05 	&		0.01 $\pm$ 0.11\\
$\mathrm{[Ca/Fe]}$ & [--0.3,--0.1] & 0.06 $\pm$ 0.05	&		0.01 $\pm$ 0.04 \\
$\mathrm{[Ca/Fe]}$ & [--0.1, 0.2]  & --0.02 $\pm$ 0.03	&	 --0.03 $\pm$ 0.03 \\
\enddata
\end{deluxetable}

From a comparisons to other abundances we suggest the [Fe/H] dependence on effective temperature is due to selection effects; cooler M giants are more likely to be metal-rich. However, the coolest stars, at temperatures of $\sim$ 3500 K and cooler, are offset in [Ti/Fe] and [Ca/Fe] from typical disk abundances. Possible problems with the atmospheric models may be affecting the [Fe/H] derivations in the coolest stars in the sample. The interpolation code for the MARCS models was not used for stars at 3500 K and cooler and deviations from measured atmospheric parameters to those adopted with the sparser grid of models may be causing the offsets. When interpreting the data, we will only analyze stars with $T_{\mathrm{eff}}$ $>$ 3500 K, and distinguish abundances from warmer and cooler stars in figures. 

\begin{deluxetable}{cccc}
\tablewidth{0pt} 
\tabletypesize{\small}
\tablecaption{Average Ti Abundances \label{table::average_ti}} 
 \tablehead{\colhead{[Fe/H] Bins} & \colhead{This Work} & \colhead{This Work} & \colhead{Bensby+14}\\
 \colhead{} & \colhead{Full Sample} & \colhead{$T_{\mathrm{eff}}$ $>$ 3500 K} & \colhead{} \\ \colhead{} & \colhead{$<$[Ti/Fe]$>$} & \colhead{$<$[Ti/Fe]$>$} & \colhead{$<$[Ti/Fe]$>$} }
\startdata
[--0.1, 0.25]  & --0.11 & --0.05 & 0.02 \\
$\mathrm{[}$--0.3, --0.1$\mathrm{]}$ & 0.05 & 0.06 & 0.08 \\
$\mathrm{[}$--0.6, --0.3$\mathrm{]}$ & 0.13 & 0.20 & 0.17\\
\enddata
\end{deluxetable}

\section{Discussion}
\label{sec:discussion}
\subsection{Cl Evolution in the Galaxy}
\begin{figure}[tp!]
\centering
\includegraphics[trim=0cm 0cm 0cm 0cm, scale=.4]{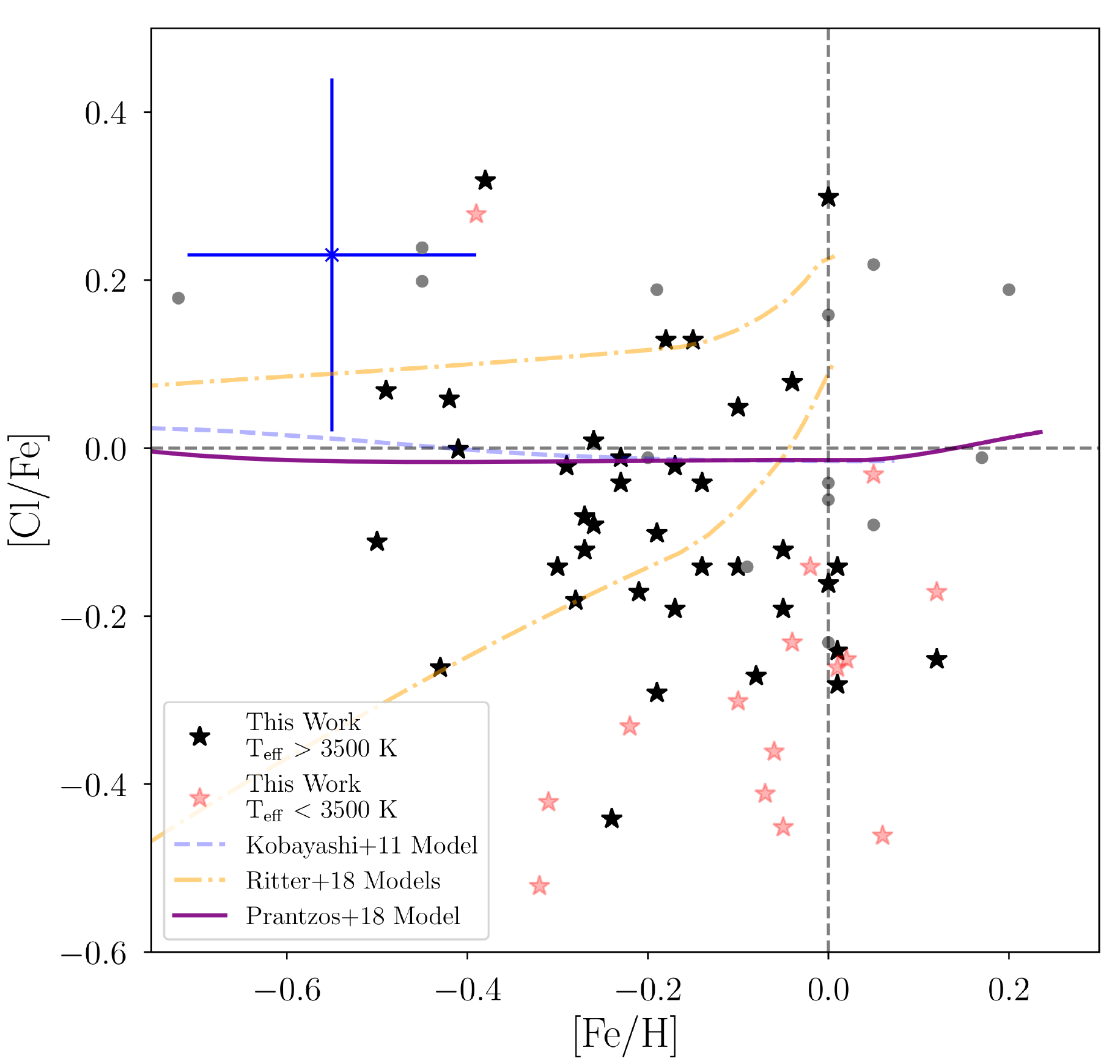}
\caption{[Cl/Fe] abundances versus [Fe/H] for M giant stars. The black stars denote abundances in stars with $T_{\mathrm{eff}}$ $>$ 3500 K and red stars denote stars with $T_{\mathrm{eff}}$ $<$ 3500 K, for stars listed in Table \ref{table::mgiant_results}. Grey circles are for [Cl/Fe] abundances from \citet{maas16}. The [Cl/Fe] ratio was calculated with $^{35}$Cl abundances with an isotope ratio of $^{35}$Cl/$^{37}$Cl = 2.66 assumed from \citet{maas18}. Chemical evolution models from \citet{kobayashi11} (blue dashed line), \citet{prantzos18} (purple solid line), and \citet{ritter18} (orange dot-dashed lines) are included. The model from \citet{ritter18} with larger [Cl/Fe] ratio assumes a C-O shell merger rate of 10$\%$ and the other model assumes no stars undergo a C-O shell merger. Dashed black lines are solar abundances. \label{fig:cl_chem_evol_iron}  }
\end{figure}

The Cl abundances from this work can be found in Fig. \ref{fig:cl_chem_evol_iron}. The [Cl/Fe] abundances were calculated using the A($^{35}$Cl) abundances from Table \ref{table::mgiant_results} and assuming a $^{35}$Cl/$^{37}$Cl = 2.66 isotope ratio, the average from the sample of \citet{maas18}. Though the Cl isotope ratio may vary in our sample, the uncertainty from the isotope ratio is less significant than the uncertainty from the effective temperature. For example, for an A($^{35}$Cl) = 5.13, the A(Cl) for  $^{35}$Cl/$^{37}$Cl = 2.08, 2.66, and 3.24, is 5.30, 5.27, and 5.25; while the uncertainty from varying $T_{\mathrm{eff}}$ $\pm$ 100 is 0.20 dex. 

Multiple Cl chemical evolution models are compared to our observed [Cl/Fe] ratios in Fig. \ref{fig:cl_chem_evol_iron}. A model from \citet{kobayashi11} includes hypernovae, CCSNe, Type Ia SN, and AGB star yields and is tuned to match the solar neighborhood. A similar model was created by \citet{prantzos18} with one key difference; the inclusion of yields from rotating massive stars. Finally, \citet{ritter18} constructed chemical evolution models with the yields of enhanced from C-O shell mergers for a certain fraction of the stars in the Galaxy. We plot the only models from \citet{ritter18} that are close to the measured [Cl/Fe] values in the data, the chemical evolution models with 0$\%$ of stars undergoing C-O shell mergers and the 10$\%$ model, in Fig. \ref{fig:cl_chem_evol_iron}. All chemical evolution models plotted initially assumed A(Cl)$|_{\odot}$ = 5.50, which we adjusted to match the value of A(Cl)$|_{\odot}$ = 5.25 from \citet{lodders09} and adopted for our data.  We note the chemical evolution models compared to our sample were tuned for the solar neighborhood which approximately corresponds to the sample of stars collected. The most distant star in our sample is HD 132917 at a distance of 836 pc and the median distance is 367 pc.

From initial inspection, we note the chemical evolution models from \citet{kobayashi11} and \citet{prantzos18} broadly agree with [Cl/Fe] ratios in our sample, and may slightly over-produce Cl in the solar neighborhood. Both models predict a nearly constant [Cl/Fe] vs. [Fe/H], with a slope of $\sim$ 0 for an [Cl/Fe] $\sim$ --0.02 dex over the range --0.6 $<$ [Fe/H] $<$ 0.3. To test if our abundances match this value, we took the average of our sample and the averages for the three metallicity bins discussed in subsection \ref{subsec:mgiant_systematic_atmo_params}. We only compute averages and compare results of the $T_{\mathrm{eff}}$ $>$ 3500 K sample, as the cooler stars appears offset at all metallicities relative to the rest of the sample. As discussed in subsection \ref{subsec:mgiant_systematic_atmo_params}, the stars with $T_{\mathrm{eff}}$ $<$ 3500 K may not have reliable abundances and are therefore excluded from our statistical analysis. 

The average abundance for all warm stars is $<$[Cl/Fe]$>$ = --0.08 $\pm$ 0.16 (s.t.dev), consistent with chemical evolution models. The average abundance ratios from each bin are given in Table \ref{table::average_cl} and we find all bins and the total sample averages are consistent with predictions and within one standard deviation of the model [Cl/Fe] ratio. Finally, we note that no model from \citet{ritter18} is able to match the slope or [Cl/Fe] ratio over the [Fe/H] range of our data. 

The [Cl/Fe] ratio also appears to decrease with increasing metallicity from inspection of Fig. \ref{fig:cl_chem_evol_iron} and from the binned averages in Table \ref{table::average_cl}. No slope for the [Cl/Fe] ratio vs. [Fe/H] ratio is expected from the chemical evolution models of \citet{kobayashi11,prantzos18}. We used a Monte Carlo simulation to determine the significance of the [Cl/Fe] vs. [Fe/H] slope. We generated random Gaussian numbers with an average value equal to the measured abundance ratio ([Cl/Fe]) with a standard deviation equal to the measurement uncertainty. We then determined a best fitting linear model using a Levenberg-Marquardt least-square method. The process was repeated 50,000 times with a new set of randomly generated data created for each iteration. We find the most probable slope is negative, in the case simulating [Cl/Fe] ratios only the average slope is --0.22 $\pm$ 0.23. The best fitting Levenberg-Marquardt least-square fit to the data yields an uncertainty on the slope with -0.22 $\pm$ 0.17, similar to the bootstrap method result. We also note the uncertainty on the x-axis flattens out the fit and drives the slope closer to zero. We fit the data using a \texttt{SIMEX} algorithm and find a slope of --0.38 $\pm$ 0.27 and a p-value of 0.16. In both cases the slope is negative but within 1-2 $\sigma$ of $\sim$0. Therefore, our abundances suggests a small slope appears to exist in the [Cl/Fe] vs. [Fe/H] data, for stars with $T_{\mathrm{eff}}$ $>$ 3500 K, but is not statistically significant given the uncertainties on each measurement. 

\begin{deluxetable}{cccc}
\tablewidth{0pt} 
\tabletypesize{\small}
\tablecaption{Average [Cl/Fe] Abundances \label{table::average_cl}} 
 \tablehead{\colhead{[Fe/H]} & \colhead{$<$[Cl/Fe]$>$} & \colhead{$\sigma$([Cl/Fe])} & \colhead{Stars}\\
 \colhead{Range} &  \colhead{} & \colhead{} & \colhead{ in Bin} }
\startdata
Full Sample & --0.08 & 0.16 & 34\\
$\mathrm{[}$--0.1, 0.25$\mathrm{]}$  & --0.13 & 0.19 & 10\\
$\mathrm{[}$--0.3, --0.1$\mathrm{]}$  & --0.09  &   0.14  &18\\
$\mathrm{[}$--0.5, --0.3$\mathrm{]}$  & 0.01   & 0.18& 6\\
\enddata
\tablecomments{All stars have $T_{\mathrm{eff}}$ $>$ 3500 K}
\end{deluxetable}

\subsection{Comparing Cl Abundances to Ca and Ti}

\begin{figure}[tp!]
\centering
\includegraphics[trim=0cm 0cm 0cm 0cm, scale=.35]{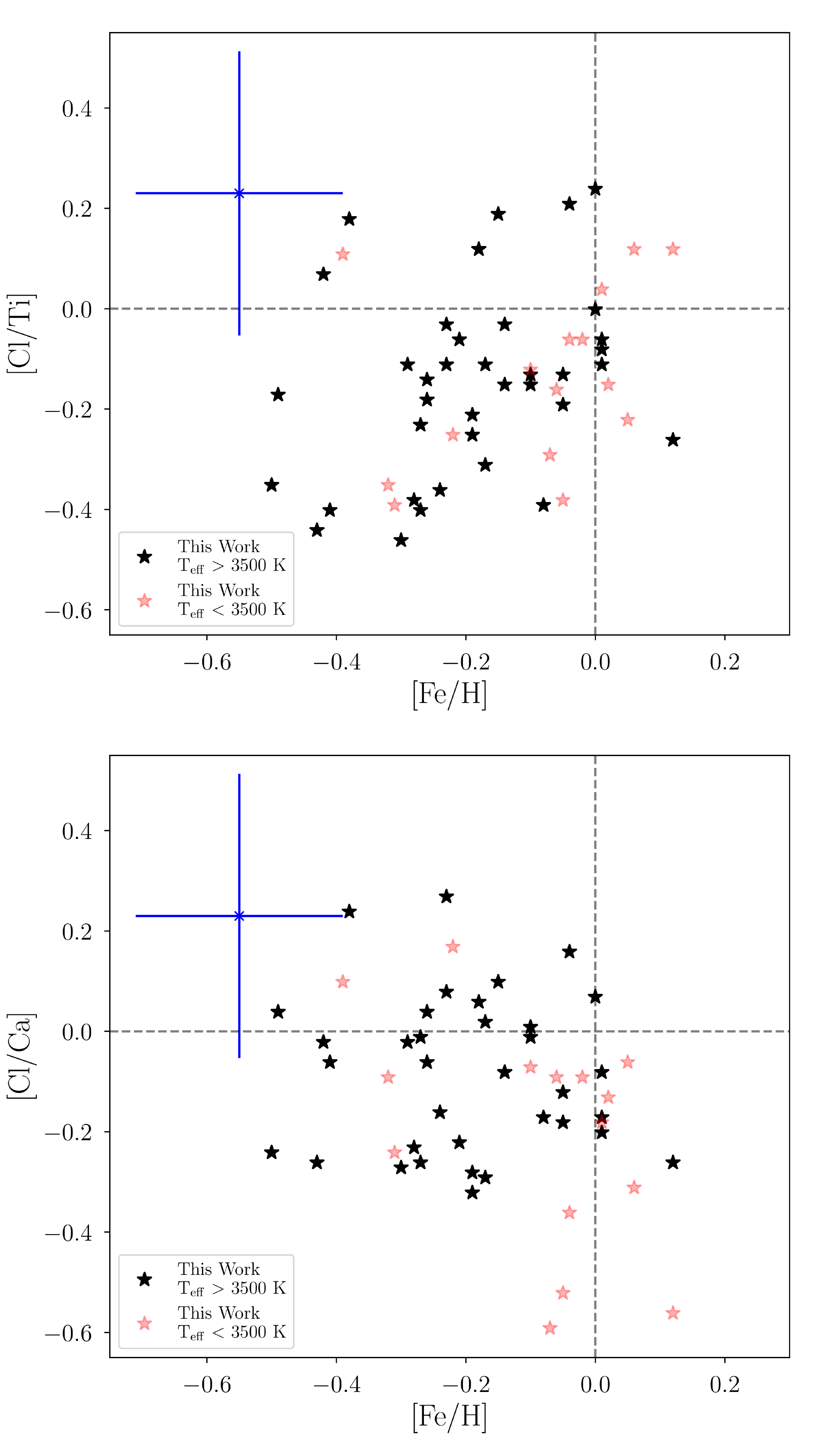}
\caption{Top Panel: [Cl/Ti] ratio versus [Fe/H]. Bottom Panel: [Cl/Ca] ratio versus [Fe/H]. The [Cl/Fe] ratio was calculated with $^{35}$Cl abundances with an isotope ratio of $^{35}$Cl/$^{37}$Cl = 2.66 assumed from \citet{maas18}. In both plots, the black stars shapes indicate stars with $T_{\mathrm{eff}}$ $>$ 3500 K and red stars indicate $T_{\mathrm{eff}}$ $<$ 3500 K. The single blue cross in each plot give representative error bars. \label{fig:cl_ca_cl_ti}  }
\end{figure}

We compare chlorine to to both Ti and Ca abundances measured in our M giants. Both elements are thought to be $\alpha$ elements produced through explosive burning in massive stars with non-trivial production from Type Ia SN. While chemical evolution models accurately reproduce Ca abundances they fail to match Ti in the solar neighborhood adding uncertainty to the CCSNe to Type Ia fraction for Ti \citep{nomoto13,prantzos18}. We note that Ca shows a shallow decrease in [Ca/Fe] as [Fe/H] as shown in Fig. \ref{fig:ti_ca_bensby}. We examine [Cl/Ca] and [Ca/Ti] in Fig. \ref{fig:cl_ca_cl_ti}. While scatter is significant, there is no slope in the [Cl/Ca] abundance ratios and the [Cl/Ti] ratio increases with [Fe/H]. The [Cl/Ca] ratio in particular indicates Cl is made in similar CCSNe to Type Ia SN ratios as Ca over our metallicity range. 

\begin{figure}[tp!]
\centering
\includegraphics[trim=0cm 0cm 0cm 0cm, scale=.26]{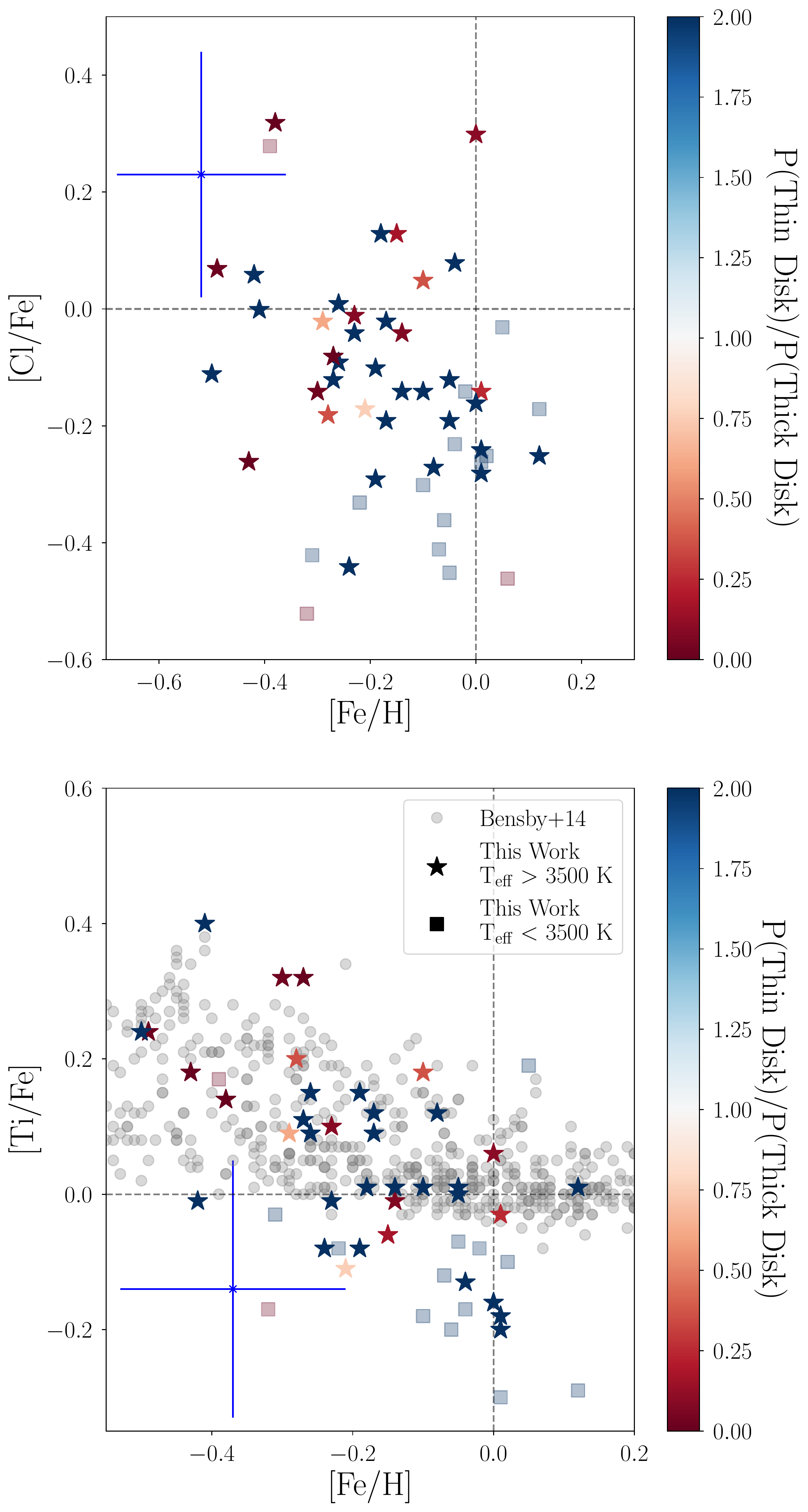}
\caption{Top Panel: [Cl/Fe] abundance versus [Fe/H] for M giant stars.  The [Cl/Fe] ratio was calculated with $^{35}$Cl abundances with an isotope ratio of $^{35}$Cl/$^{37}$Cl = 2.66 assumed from \citet{maas18}. Bottom Panel: [Ti/Fe] ratios from our sample are compared to abundances from \citet{bensby14} (grey points). In both plots, the dashed black lines are solar abundances and the stars shapes represent stars with $T_{\mathrm{eff}}$ $>$ 3500 K and squares are $T_{\mathrm{eff}}$ $<$ 3500 K. The single blue cross in each plot give representative error bars. Thin to thick disk probability ratios are given in the color bar. \label{fig:cl_thin_thick_disk}  }
\end{figure}

 We also calculated the UVW velocity of our sample using Gaia distances and proper motions \citep{gaia18}. Radial velocity measurements were adopted primarily from \citet{famaey05}. The Gaia data, with RA, DEC, and radial velocity measurements were accessed using the SIMBAD database. The UVW velocities were calculated using a python based calculator\footnote{https://github.com/dr-rodriguez/UVW\_Calculator}. The thin disk, thick disk, and halo membership probabilities based on the kinematics of each star were calculated using the same methodology as \citet{ramirez13}. The thin/thick disk probabilities with Cl and Ti abundances are compared in Fig. \ref{fig:cl_thin_thick_disk} and the probabilities are given in Table \ref{table::mgiant_uvw}. 
 
 \begin{deluxetable*}{ccccccccc}
\tablewidth{0pt} 
\tabletypesize{\small}
\tablecaption{M giant Fundamental Parameters and UVW Velocities \label{table::mgiant_uvw}} 
 \tablehead{\colhead{HD} & \colhead{Lum.} & \colhead{$\delta$Lum.} & \colhead{Distance\tablenotemark{a}} & \colhead{Mass}   & \colhead{U} & \colhead{V} & \colhead{W} & \colhead{P(Thin/Thick)} \\ \colhead{Num} &  \colhead{(L$_{\odot}$)}& \colhead{(L$_{\odot}$)} & \colhead{(pc.)} & \colhead{(M$_{\odot}$)} &\colhead{(km s$^{-1}$)}& \colhead{(km s$^{-1}$)}& \colhead{(km s$^{-1}$)}& \colhead{Ratio} }
\startdata
5111	&	530	&	170   &	351	& 	1	&	84.5	&	-69.6	&	-12.7	&	0.3	\\
15594	&	2040	&	670	&	383	& 	1.6	&	-58.6	&	8.5	&	16.0	&	51.5	\\
19258	&	920	&	190   &	349	& 	1.4	&	55.9	&	-49.0	&	48.6	&	0.8	\\
25000	&	1600	&	440	&	372	& 	1.6	&	47.4	&	-33.1	&	-10.5	&	25.3	\\
30634	&	810	&	170 	&	336	& 	1.5	&	10.6	&	-25.1	&	-49.7	&	7.0	\\
\enddata
\tablenotetext{a}{Distances from \citet{bailer18}}
\tablecomments{(This table is available in its entirety in machine-readable form.)}
\end{deluxetable*}

 We identify 17 stars in our sample as probable thick disk members with probability ratios of P$_{thin}$/P$_{thick}$ $<$ 1. The average iron abundance for thin and thick disk stars  with $T_{\mathrm{eff}}$ $>$ 3500 K are $<$[Fe/H]$>$ = --0.17 $\pm$ 0.15 and $<$[Fe/H]$>$ = --0.23 $\pm$ 0.15 respectively. In the top panel of Fig. \ref{fig:cl_thin_thick_disk} we note that the thin and thick disk stars have similar [Ti/Fe] ratios at [Fe/H] $>$ --0.20 dex. Beyond this metallicity the [Ti/Fe] ratios are larger than the [Fe/H] in Fig. \ref{fig:cl_thin_thick_disk}. For Cl, the difference between [Cl/Fe] abundances between thin disk and thick disk members at any [Fe/H] is not significant. The consistency between both thin and thick disk populations suggests that Cl production is significant enough in Type Ia SN to cause no detectable difference in either population for our sample, although our sample sizes are small and uncertainties relatively large.  

\subsection{Stellar Cl Abundances Compared to Nebular Cl Abundances}

\begin{figure}[tp!]
	\centering
	\includegraphics[trim=0cm 0cm 0cm 0cm, scale=.32]{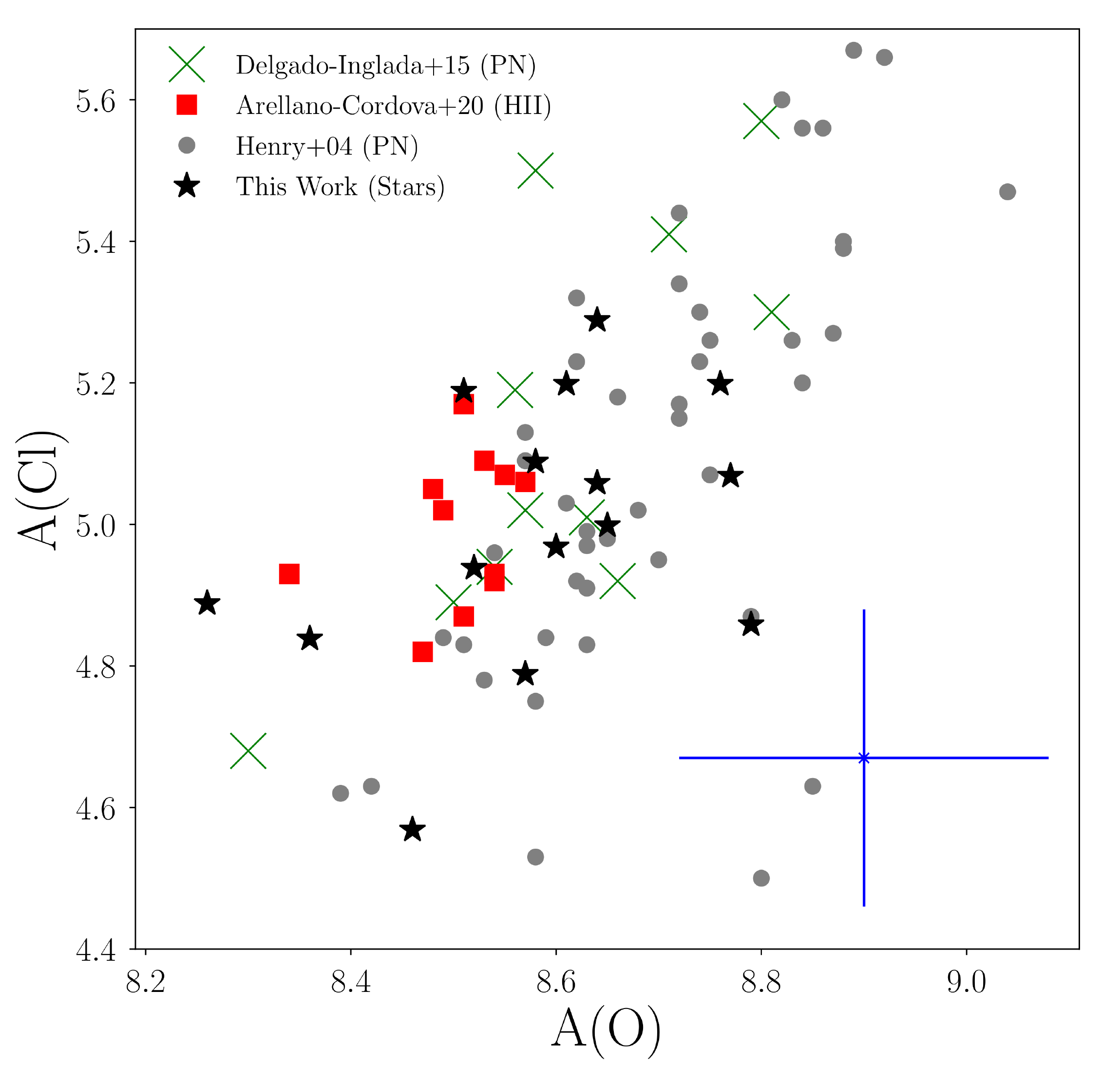}
	\caption{A(Cl) vs. A(O) for nebular abundance data and from our sample of stars. Red squares are from \citet{arellano20}, orange x shapes are from \citet{delgado15}, grey circles are from \citet{henry04}, blue diamonds are from \citet{maas16}, and black stars are from this work. The A(Cl) abundances use the A($^{35}$Cl) abundances and adopt an isotope ratio of $^{35}$Cl/$^{37}$Cl = 2.66 from \citet{maas18} to calculate total Cl abundances.  \label{fig:cl_h2_pn}  }
	\end{figure}

 We compare the derived O and Cl abundances in stars to measurements in H II regions and PNs in Fig. \ref{fig:cl_h2_pn}. We select H II and PN nebular abundances from \citet{henry04,delgado15,arellano20}. Since all of our target stars have distances within 1 kpc from the Sun, we select objects with  galactocentric radii between 6 kpc and 11 kpc. The cut ensures we are not comparing the chemical evolution of different regions of the Galaxy to our sample. The nebular and stellar A(O) and A(Cl) abundances are plotted in Fig. \ref{fig:cl_h2_pn}.
 
We find the stellar abundances broadly agree with the H II region and PN A(Cl) measurements, although there is significant scatter in the data. The consistent results suggests both sets of astrophysical objects accurately probe the Cl abundance, providing evidence no significant systematic uncertainties affect the Cl abundances in stars measured with HCl features. We also find A(Cl) and A(O) increasing in all samples, suggesting both A(Cl) and A(O) are produced in massive stars and are made in lockstep, providing additional evidence the primary source of Cl is from CCSNe. We note the scatter and uncertainty on the Cl abundances are large, making a search for secondary behavior with respect to oxygen difficult. A simple linear regression for all objects in Fig. \ref{fig:cl_h2_pn}, excluding objects with A(Cl) $<$ 4.75 to remove outliers, finds a slope of 1.03 $\pm$ 1.01. More precise measurements in more stars are needed to further constrain the evolution of Cl but the general agreement between nebular abundances and stellar abundances are promising for the accuracy of stellar Cl abundance measurements. 
 
\section{Conclusions}
\label{sec:conclusion}
 The Galactic chemical evolution and nucleosynthesis of Cl was examined a sample of 52 M giants. Atmospheric parameters were derived using optical echelle spectra of our M giants with $T_{\mathrm{eff}}$ from \citet{mcdonald12}, log(g) using Gaia DR2 parallax measurements, [Fe/H] and microturbulence from Fe I equivalent widths. We then measured $^{35}$Cl, Ca, and Ti abundances using MOOG and MARCS atmospheric models. We tested for systematic uncertainties from our atmospheric parameters and excluded stars with $T_{\mathrm{eff}}$ $<$ 3500 K. Our results on the Galactic chemical evolution and nucleosynthesis of Cl from this data set are given below:
 
\begin{enumerate}
 
\item{From our stellar sample, we find the binned average [Cl/Fe] ratios broadly agree with chemical evolution models from \citet{kobayashi11,prantzos18} but do not match the chemical evolution model of \citet{ritter18}. Agreement between the average abundances and models provides evidence that Cl yields in CCSNe, Type Ia SN, and AGB stars, with Cl primarily produced in CCSNe, are sufficient to successfully predict Cl abundances in the solar neighborhood. We suggest Cl is not significantly produced through O-C shell mergers or by the $\nu$ process over the metallicity range of our sample.}
 
\item{A small slope observed in the [Cl/Fe] vs. [Fe/H] relationship in Fig. \ref{fig:cl_chem_evol_iron} was not predicted by chemical evolution models. We performed a Monte Carlo experiment to determine the significance of the slope with the results. We find the slope in the data is not statistically significant due to the large uncertainties on our measured abundance ratios. The difficulty is due to the large uncertainties on the $^{35}$Cl abundances, primarily from uncertainty on $T_{\mathrm{eff}}$. More precise $T_{\mathrm{eff}}$ measurements would be useful in future studies of Cl in M giants.}

\item{Comparisons of Cl to Ca and Ti found that the [Cl/Ca] ratio is approximately constant over the observed [Fe/H] range of the sample while [Cl/Ti] increases with  increasing [Fe/H]. These comparisons suggest Cl is made in similar nucleosynthesis sites as Ca; a mix of mostly CCSNe contributions with a significant Type Ia SN contribution. A similar mix of Cl yields at both nucleosynthesis sites also explains why no significant difference is observed between Cl abundances in the thin and thick disk stars in our sample. }
 
\item{Chlorine and oxygen abundances in M giants are consistent with measurements made in planetary nebulae \citep{henry04, delgado15} and \ion{H}{2} regions \citep{arellano20}, although there is significant scatter in the data.}

\end{enumerate}

\section*{Acknowledgements}
This work is based on observations obtained with the Apache Point Observatory 3.5-meter telescope, which is owned and operated by the Astrophysical Research Consortium. Visiting Astronomer at the Infrared Telescope Facility, which is operated by the University of Hawaii under contract NNH14CK55B with the National Aeronautics and Space Administration. The authors wish to recognize and acknowledge the very significant cultural role and reverence that the summit of Mauna Kea has always had within the indigenous Hawaiian community.  We are most fortunate to have the opportunity to conduct observations from this mountain. This research has made use of the NASA Astrophysics Data System Bibliographic Services and the Kurucz atomic line database operated by the Center for Astrophysics. This research has made use of the SIMBAD database, operated at CDS, Strasbourg, France. This publication makes use of data products from the Two Micron All Sky Survey, which is a joint project of the University of Massachusetts and the Infrared Processing and Analysis Center/California Institute of Technology, funded by the National Aeronautics and Space Administration and the National Science Foundation. This publication makes use of data products from the Wide-field Infrared Survey Explorer, which is a joint project of the University of California, Los Angeles, and the Jet Propulsion Laboratory/California Institute of Technology, funded by the National Aeronautics and Space Administration. We thank the anonymous referee for their thoughtful comments and suggestions on the manuscript. We thank Eric Ost for implementing the model atmosphere interpolation code. C. A. P. acknowledges the generosity of the Kirkwood Research Fund at Indiana University. 

\software{IRAF \citep{tody86,tody93}, PyRAF \citep{pyraf12}, MOOG (v2017; \citealt{sneden73}), \texttt{scipy} \citep{jones01}, \texttt{numpy} \citep{walt11}, \texttt{matplotlib} \citep{hunter07}, \texttt{UVW calculator} (\url{https://github.com/dr-rodriguez/UVW_Calculator})}


\begin{thebibliography}{}

\bibitem[Arellano-C{\'o}rdova et al.(2020)]{arellano20} 
Arellano-C{\'o}rdova, K.~Z., Esteban, C., Garc{\'\i}a-Rojas, J., et al.\ 2020, \mnras, 496, 1051

\bibitem[Asplund et al.(2009)]{asplund09} 
Asplund, M., Grevesse, N., Sauval, A.~J., et al.\ 2009, \araa, 47, 481

\bibitem[Bailer-Jones et al.(2018)]{bailer18} 
Bailer-Jones, C.~A.~L., Rybizki, J., Fouesneau, M., et al.\ 2018, \aj, 156, 58

\bibitem[Bensby et al.(2014)]{bensby14} 
Bensby, T., Feltzing, S., \& Oey, M.~S.\ 2014, \aap, 562, A71

\bibitem[Bertelli et al.(2008)]{bertelli08} 
Bertelli, G., Girardi, L., Marigo, P., et al.\ 2008, \aap, 484, 815

\bibitem[Bertelli et al.(2009)]{bertelli09} 
Bertelli, G., Nasi, E., Girardi, L., et al.\ 2009, \aap, 508, 355

\bibitem[Bessell et al.(1998)]{bessell98} 
Bessell, M.~S., Castelli, F., \& Plez, B.\ 1998, \aap, 333, 231

\bibitem[Cernicharo et al.(2000)]{cernicharo00} 
Cernicharo, J., Gu{\'e}lin, M., \& Kahane, C.\ 2000, \aaps, 142, 181


\bibitem[Cernicharo et al.(2010)]{cernicharo10} 
Cernicharo, J., Goicoechea, J.~R., Daniel, F., et al.\ 2010, \aap, 518, L115

\bibitem[Chou et al.(2007)]{chou07} 
Chou, M.-Y., Majewski, S.~R., Cunha, K., et al.\ 2007, \apj, 670, 346

\bibitem[Cristallo et al.(2015)]{cristallo15}
Cristallo, S., Straniero, O., Piersanti, L., \& Gobrecht, D.\ 2015, \apjs, 219, 40

\bibitem[Delgado-Inglada et al.(2015)]{delgado15}
Delgado-Inglada, G., Rodr{\'{\i}}guez, M., Peimbert, M., Stasi{\'n}ska, G., 
\& Morisset, C.\ 2015, \mnras, 449, 1797

\bibitem[ESA(1997)]{esa97} 
ESA\ 1997, ESA Special Publication

\bibitem[Esteban et al.(2015)]{esteban15} 
Esteban, C., Garc{\'{\i}}a-Rojas, J., \& P{\'e}rez-Mesa, V.\ 2015, \mnras, 452, 1553 

\bibitem[Famaey et al.(2005)]{famaey05} 
Famaey, B., Jorissen, A., Luri, X., et al.\ 2005, \aap, 430, 165

\bibitem[Fang et al.(2018)]{fang18} 
Fang, X., Garc{\'\i}a-Benito, R., Guerrero, M.~A., et al.\ 2018, \apj, 853, 50

\bibitem[Gaia Collaboration et al.(2018)]{gaia18} 
Gaia Collaboration, Brown, A.~G.~A., Vallenari, A., et al.\ 2018, \aap, 616, A1

\bibitem[Ge et al.(2016)]{ge16} Ge, J., Ma, B., Sithajan, S., et al.\ 2016, \procspie, 99086I

\bibitem[Greene et al.(1993)]{greene93} 
Greene, T.~P., Tokunaga, A.~T., Toomey, D.~W., et al.\ 1993, \procspie, 313

\bibitem[Gustafsson et al.(2008)]{gustafsson08} Gustafsson, B., Edvardsson, B., Eriksson, K., et al.\ 2008, \aap, 486, 951

\bibitem[Hall \& Noyes(1972)]{hall72} 
Hall, D.~N.~B. \& Noyes, R.~W.\ 1972, \apjl, 175, L95

\bibitem[Hauser et al.(1998)]{hauser98}
Hauser, M. G., Kelsall, T., Leisawitz, D., $\&$ Weiland, J. 1998,COBE Diffuse Infrared Background Experiment (DIRBE) Explanatory Supplement,vers. 2.3 (Greenbelt: NASA)

\bibitem[Henry et al.(2004)]{henry04} 
Henry, R.~B.~C., Kwitter, K.~B., \& Balick, B.\ 2004, \aj, 127, 2284 

\bibitem[Highberger et al.(2003)]{highberger03} 
Highberger, J.~L., Thomson, K.~J., Young, P.~A., Arnett, D., \& Ziurys, L.~M.\ 2003, \apj, 593, 393

\bibitem[Hinkle et al.(2000)]{hinkle00} Hinkle, K., Wallace, L., Valenti, J., et al.\ 2000, Visible and Near Infrared Atlas of the Arcturus Spectrum 3727-9300 A ed. Kenneth Hinkle

\bibitem[Hunter(2007)]{hunter07}
Hunter, J. D. 2007, Computing in Science $\&$ Engineering, 9,
90. \url{http://scitation.aip.org/content/aip/journal/
cise/9/3/10.1109/MCSE.2007.55}

\bibitem[Jian et al.(2017)]{jian17} 
Jian, M., Gao, S., Zhao, H., et al.\ 2017, \aj, 153, 5

\bibitem[Johnson et al.(2005)]{johnson05} 
Johnson, C.~I., Kraft, R.~P., Pilachowski, C.~A., et al.\ 2005, \pasp, 117, 1308

\bibitem[Jones et al.(2001)]{jones01} 
Jones, E., Oliphant, T., Peterson, P., et al. 2001, SciPy: 
Open source scientific tools for Python, , .\url{http://www.scipy.org/s}

\bibitem[Joyce(1992)]{joyce92}
Joyce, R. R. 1992, in ASP Conf. Ser. 23, Astronomical CCD Observing and
Reduction Techniques, ed. S. Howell (San Francisco: ASP), 258

\bibitem[Kahane et al.(2000)]{kahane00}
Kahane, C., Dufour, E., Busso, M., et al.\ 2000, \aap, 357, 669

\bibitem[Kama et al.(2015)]{kama15} 
Kama, M., Caux, E., L{\'o}pez-Sepulcre, A., et al.\ 2015, \aap, 574, A107

\bibitem[Karakas \& Lugaro(2016)]{karakas16}
Karakas, A.~I., \& Lugaro, M.\ 2016, \apj, 825, 26 

\bibitem[Kobayashi et al.(2011)]{kobayashi11} Kobayashi, C., Karakas, A.~I., \& Umeda, H.\ 2011, \mnras, 414, 3231 

\bibitem[Kobayashi et al.(2020)]{kobayashi20} 
Kobayashi, C., Karakas, A.~I., \& Lugaro, M.\ 2020, \apj, 900, 179


\bibitem[Leung \& Nomoto(2018)]{leung18} 
Leung, S.-C., \& Nomoto, K.\ 2018, \apj, 861, 143

\bibitem[Lodders et al.(2009)]{lodders09} 
Lodders, K., Palme, H., \& Gail, H.-P.\ 2009, Landolt B{\"o}rnstein, 

\bibitem[Maas et al.(2016)]{maas16} 
Maas, Z.~G., Pilachowski, C.~A., \& Hinkle, K.\ 2016, \aj, 152, 196

\bibitem[Maas \& Pilachowski(2018)]{maas18} 
Maas, Z.~G., \& Pilachowski, C.~A.\ 2018, \aj, 156, 2

\bibitem[Maas(2020)]{maas20} 
Maas, Zachary G., The Galactic Chemical Evolution of Chlorine and Phosphorus, Dissertation, Indiana University, Ann Arbor, 2020. ProQuest

\bibitem[Martin $\&$ Hepburn (1998)]{martin98}
Martin J. D. D.  $\&$  Hepburn J. W., 1998, J. Phys. Chem., 109, No. 15, 8139

\bibitem[McDonald et al.(2012)]{mcdonald12} 
McDonald, I., Zijlstra, A.~A., \& Boyer, M.~L.\ 2012, \mnras, 427, 343

\bibitem[Milingo et al.(2010)]{milingo10} 
Milingo, J.~B., Kwitter, K.~B., Henry, R.~B.~C., et al.\ 2010, \apj, 711, 619

\bibitem[Muller et al.(2014)]{muller14}
 Muller, S., Black, J.~H., Gu{\'e}lin, M., et al.\ 2014, \aap, 566, L6

\bibitem[Neufeld et al.(2012)]{neufeld12}
 Neufeld, D.~A., Roueff, E., Snell, R.~L., et al.\ 2012, \apj, 748, 37 

\bibitem[Neufeld et al.(2015)]{neufeld15} 
Neufeld, D.~A., Black, J.~H., Gerin, M., et al.\ 2015, \apj, 807, 54

\bibitem[Nomoto et al.(2013)]{nomoto13}
Nomoto, K., Kobayashi, C., \& Tominaga, N.\ 2013, \araa, 51, 457 

\bibitem[Peng et al.(2010)]{peng10} 
Peng, R., Yoshida, H., Chamberlin, R.~A., et al.\ 2010, \apj, 723, 218

\bibitem[Pignatari et al.(2010)]{pignatari10} 
Pignatari, M., Gallino, R., Heil, M., et al.\ 2010, \apj, 710, 1557

\bibitem[Prantzos et al.(1990)]{prantzos90} 
Prantzos, N., Hashimoto, M., \& Nomoto, K.\ 1990, \aap, 234, 211

\bibitem[Prantzos et al.(2018)]{prantzos18}
Prantzos, N., Abia, C., Limongi, M., et al.\ 2018, \mnras, 476, 3432

\bibitem[Plez(1998)]{plez98} 
Plez, B.\ 1998, \aap, 337, 495

\bibitem[Ram{\'\i}rez et al.(2013)]{ramirez13} 
Ram{\'\i}rez, I., Allende Prieto, C., \& Lambert, D.~L.\ 2013, \apj, 764, 78

\bibitem[Ram{\'\i}rez, \& Allende Prieto(2011)]{ramirez11} 
Ram{\'\i}rez, I., \& Allende Prieto, C.\ 2011, \apj, 743, 135

\bibitem[Ram{\'\i}rez, \& Mel{\'e}ndez(2005)]{ramirez05} 
Ram{\'\i}rez, I., \& Mel{\'e}ndez, J.\ 2005, \apj, 626, 465

\bibitem[Ritter et al.(2018)]{ritter18} 
Ritter, C., Andrassy, R., C{\^o}t{\'e}, B., et al.\ 2018, \mnras, 474, L1 

\bibitem[Ruland et al.(1980)]{ruland80} 
Ruland, F., Holweger, H., Griffin, R., et al.\ 1980, \aap, 92, 70

 \bibitem[Salez et al.(1996)]{salez96} 
 Salez, M., Frerking, M.~A., \& Langer, W.~D.\ 1996, \apj, 467, 708

\bibitem[Schlafly \& Finkbeiner(2011)]{schlafly11} 
Schlafly, E.~F., \& Finkbeiner, D.~P.\ 2011, \apj, 737, 103

\bibitem[Science Software Branch at STScI(2012)]{pyraf12} 
Science Software Branch at STScI\ 2012, Astrophysics Source Code Library. ascl:1207.011

\bibitem[Skrutskie et al.(2006)]{skrutskie06} 
Skrutskie, M.~F., Cutri, R.~M., Stiening, R., et al.\ 2006, \aj, 131, 1163

\bibitem[Smith et al.(2004)]{smith04} 
Smith, B.~J., Price, S.~D., \& Baker, R.~I.\ 2004, \apjs, 154, 673

\bibitem[Smith, \& Lambert(1990)]{smith90}
Smith, V.~V., \& Lambert, D.~L.\ 1990, \apjs, 72, 387

\bibitem[Sneden(1973)]{sneden73} 
Sneden, C.\ 1973, \apj, 184, 839

\bibitem[Thielemann \& Arnett(1985)]{thielemann85} 
Thielemann, F.~K., \& Arnett, W.~D.\ 1985, \apj, 295, 604

\bibitem[Tody(1993)]{tody93} 
Tody, D.\ 1993, Astronomical Data Analysis Software and Systems II, 173

\bibitem[Tody(1986)]{tody86} 
Tody, D.\ 1986, \procspie, 733

\bibitem[Tokunaga et al.(1990)]{tokunaga90} 
Tokunaga, A.~T., Toomey, D.~W., Carr, J., et al.\ 1990, \procspie, 131

\bibitem[Tomkin, \& Lambert(1983)]{tomkin83} 
Tomkin, J., \& Lambert, D.~L.\ 1983, \apj, 273, 722

\bibitem[Travaglio et al.(2004)]{travaglio04} 
Travaglio, C., Hillebrandt, W., Reinecke, M., \& Thielemann, F.-K.\ 2004, \aap, 425, 1029

\bibitem[van der Walt et al.(2011)]{walt11}
van der Walt, S., Colbert, S. C., $\&$ Varoquaux, G. 2011,
Computing in Science $\&$ Engineering, 13, 22.
\url{http://scitation.aip.org/content/aip/journal/cise/13/2/10.1109/MCSE.2011.37}

\bibitem[Wallace et al.(2011)]{wallace11} 
Wallace, L., Hinkle, K.~H., Livingston, W.~C., et al.\ 2011, The Astrophysical Journal Supplement Series, 195, 6

\bibitem[Wallstr{\"o}m et al.(2019)]{wallstrom19} 
Wallstr{\"o}m, S.~H.~J., Muller, S., Roueff, E., et al.\ 2019, \aap, 629, A128

\bibitem[Wang et al.(2003)]{wang03} 
Wang, S.-. i ., Hildebrand, R.~H., Hobbs, L.~M., et al.\ 2003, \procspie, 1145

\bibitem[Woosley et al.(1973)]{woosley73} 
Woosley, S.~E., Arnett, W.~D., \& Clayton, D.~D.\ 1973, \apjs, 26, 231

\bibitem[Woosley et al.(1990)]{woosley90} 
Woosley, S.~E., Hartmann, D.~H., Hoffman, R.~D., et al.\ 1990, \apj, 356, 272

\bibitem[Woosley \& Weaver(1995)]{woosley95}
Woosley, S.~E., \& Weaver, T.~A.\ 1995, \apjs, 101, 181 

\bibitem[Wright et al.(2010)]{wright10} Wright, E.~L., Eisenhardt, P.~R.~M., Mainzer, A.~K., et al.\ 2010, \aj, 140, 1868

\end{thebibliography}
\end{document}